\documentclass[11pt,review,1p,number,sort&compress,a4paper]{elsarticle}
\usepackage{dcolumn}

\usepackage[utf8]{inputenc}
\usepackage{graphics,graphicx}
\usepackage{epstopdf}
\usepackage{float}
\usepackage{natbib}
\usepackage{fancyvrb}
\usepackage{subfigure}
\usepackage{hyperref}
\usepackage{pifont}
\usepackage[version=4]{mhchem}

\usepackage{textcomp}
\usepackage{upquote}

\usepackage{color}
\usepackage{longtable}
\usepackage{dcolumn}
\newcolumntype{d}[1]{D{.}{.}{#1}}

\newcommand{\tick}{\ding{52}}

\newcommand{\ea}{{\it et al}}
\newcommand{\cm}{cm$^{-1}$}

\newcommand{\Cv}[1]{${\mathcal C}_{#1{\rm v}}$}

\newcommand{\pp}{^{\prime\prime}}

\journal{J. Quant. Spectrosc. Radiat. Transf.}

\begin{document}

\begin{frontmatter}

\title{The 2020 release of the ExoMol database: molecular line lists for exoplanet and other hot
atmospheres}
\author{Jonathan Tennyson$^{1}$\thanks{Corresponding author}}
\ead{j.tennyson@ucl.ac.uk}
\author{Sergei  N. Yurchenko$^{1}$}
\author{Ahmed F. Al-Refaie$^{1}$}
\author{Victoria H. J. Clark$^{1}$}
\author{Katy L. Chubb$^{1,2}$}
\author{Eamon K. Conway$^{1,9}$}
\author{Akhil Dewan$^{1}$}
\author{Maire N. Gorman$^{1,3}$}
\author{Christian Hill$^{1,4}$}
\author{A. E. Lynas-Gray$^{1,7,8}$} 
\author{Thomas Mellor$^{1}$}
\author{Laura K. McKemmish$^{1,5}$}
\author{Alec Owens$^{1}$}
\author{Oleg L. Polyansky$^{1}$}
\author{Mikhail Semenov$^{1}$}
\author{Wilfrid Somogyi$^{1}$}
\author{Giovanna Tinetti$^{1}$}
\author{Apoorva Upadhyay$^{1}$}
\author{Ingo Waldmann$^{1}$}
\author{Yixin Wang$^{1,6}$}
\author{Samuel Wright$^1$}
\author{Olga P. Yurchenko$^{1}$}
\address{$^{1}$Department of Physics and Astronomy, University College London, London,
WC1E 6BT, UK\\
	$^{2}$Present address: SRON Netherlands Institute for Space Research, Sorbonnelaan 2,
        3584 CA, Utrecht, Netherlands\\
		$^{3}$ Present address: Department of Physics, Aberystwyth University, Ceredigion, UK, SY23~3BZ, UK \\
$^{4}$Present address: Nuclear Data Section, International Atomic Energy Agency, Vienna A-1400, Austria\\
	$^{5}$ Present address: School of Chemistry, University of New South Wales, 2052 Sydney, Australia\\
	$^{6}$ Present address: Nankai
University, 94 Weijin Road, Tianjin, China \\
    $^{7}$ Department of Physics, University of Oxford, England \\
    $^{8}$ Department of Physics and Astronomy, University of the Western Cape, South Africa \\
    $^{9}$ Atomic and Molecular Physics Division, Center for Astrophysics $|$ Harvard \& Smithsonian, Cambridge, MA. USA.}
\date{\today}

\begin{abstract}

  The ExoMol database (www.exomol.com) provides molecular data for
  spectroscopic studies of hot atmospheres. While the data is
  intended for studies of exoplanets and other astronomical bodies,
  the dataset is widely applicable. The basic form of the database is
  extensive line lists; these are supplemented with partition functions,
  state lifetimes, cooling functions, Land\'e g-factors, temperature-dependent cross sections, opacities, pressure broadening parameters, 
  $k$-coefficients and dipoles. This paper presents the latest release of
  the database which has been expanded to consider 80 molecules
  and 190 isotopologues totaling over 700 billion transitions. While the spectroscopic data is concentrated   at infrared and visible wavelengths, ultraviolet transitions are
  being increasingly considered in response to requests from observers.
  The core of the database comes from the ExoMol project which primarily uses
  theoretical methods, albeit usually fine-tuned to reproduce laboratory spectra,
  to generate very extensive line lists for studies of hot bodies. The
  data has recently been supplemented by line lists deriving from direct
  laboratory observations, albeit usually with the use of {\it ab initio}
  transition intensities. A major push in the new release is towards accurate
  characterisation of transition frequencies for use in high resolution
  studies of exoplanets and other bodies.
\end{abstract}

\end{frontmatter}

\newpage

\section{Introduction}

The ExoMol project was started in 2011 \cite{jt528} with the purpose of providing
molecular line lists for studies of exoplanets  and other (hot) atmospheres.
Besides data demands for exoplanets \cite{13TiEnCo.exo}, other hot astronomical bodies
 with significant molecular content in their atmospheres include cool stars
\cite{09Bernath.exo} and brown dwarfs
\cite{aha97,09Bernath.exo}. On Earth, similar spectroscopic data are required
to study flames
\cite{14BoWeHy.H2O,15CoLixx}, discharge plasmas \cite{02ChWaLi.SO},
explosions \cite{11CaLiPi.H2O} and the gases emitted
 from smoke stacks \cite{12EvFaCl.CO2}. In addition, 
bodies in   non-local thermodynamic equilibrium (non-LTE) such as
comets and masers give observable emissions from highly excited
states \cite{jt330,jt349,16GrBaRi.H2O}. 

ExoMol data are proving  very popular among the exoplanet atmospheric modellers owing to their extensive coverage of the molecular species and their completeness as function of both frequency  and, very importantly, temperature. ExoMol data have been incorporated into many radiative transfer and retrieval codes including Tau-REx \cite{jt593,jt611,TauRexIII},  Phoenix \cite{PHOENIX},
NEMESIS \cite{NEMESIS}, CHIMERA \cite{CHIMERA}, HELIOS \cite{HELIOS} and HELIOS-r2 \cite{Helios-r2}, ATMO \cite{ATMO,18GoMaSi.exogen,20PhTrBa},
 ARCiS \cite{19MiOrCh.arcis}, ARTES \cite{ARTES}, HyDRA \cite{HyDRA}, GENESIS \cite{17GaMaxx.exoplanet},
petitRADTRANS \cite{19MoWaBo.petitRADTRANS}, PLATON \cite{PLATON},
VSTAR \cite{12BaKe,jt572}, BART \cite{BART} and Pyrat Bay \cite{PyratBay}. 
In addition the ExoMol database has been used to support a variety of other studies. Examples including the analysis of gaseous ammonia in Jupiter \cite{jt745}, tentative
detection  of 
H$_2$S in Uranus \citep{19IrToGa.H2S}, detection of  CaO in meteors \citep{18BeBoSa.CaO} and  SiO in B[e] Supergiants \citep{15KrOkCi.SiO}, Late-type stars \citep{19Pavlenko} or  circumstellar environment \citep{19EvPaBa}. 
Detections in the atmospheres of exoplanets are discussed below.
More diverse  applications include  modeling
laser analysis of Solar System objects \cite{15HuLuMe,16HuLuCo}, isotope abundance quantification in stars \cite{jt799}, search for molecules highly sensitive to proton-to-electron mass ratio variation  \cite{18OwYuSp,20SyMoCu.applications}, 
models of molecular steering \cite{jt663,18OwYa} , 
design of THz lasers
\cite{19BuMaEl.applications}, design of  dark matter detection schemes \cite{19EsPeRa} and the design of novel propulsion systems \cite{20Markov.NH3}.

A number of diatomic molecules for which ExoMol provides extensive hot line lists have been detected in sunspots. These include SiO \cite{ 79JoPuPa.SiO}, SiH \cite{12SrAmSh.SiH}, SH \cite{79JoPaXX.SH}, BeH \cite{08ShBaRa.BeH}, VO \cite{08SrBaRa.VO}, as well as MgH, MgO, TiO, C$_2$ and CaH which have all long been observed in sunspots \cite{71Sotiro}. 
In addition the spectrum of water is well-known in sunspots \cite{jt200,jt212,jt251,jt297,06SoWiSc.H2O}; studies of water spectra in sunspots should
be aided by the new POKAZATEL water line list \cite{jt734} which  is designed for studies
at temperatures above 3000 K such as those encountered in solar umbra and penumbra.














 ExoMol line lists have been published as a series in the journal Monthly Notices
of the Royal Astronomical Society (see below) and a first data release in 2016 
\cite{jt631}, henceforth ExoMol2016, provided
full documentation for the ExoMol database which can be found at \url{www.exomol.com}.
The basic form of the database is
  extensive line lists; these are supplemented with partition functions,
  state lifetimes, cooling functions, Land\'e g-factors, temperature-dependent cross sections, opacities, pressure broadening parameters, 
  $k$-coefficients and dipoles. The 
 ExoMol website also offers an extensive bibliography database on research literature on molecules relevant for ExoMol applications. So far the bibliography database contains 6709 sources.

 Since ExoMol started there have been  major changes in exoplanet science which
 have driven further expansion and development of the ExoMol database.
 The first is the discovery of rocky exoplanets orbiting so near to their host
 star that their surfaces are likely to be molten. 
 The atmospheres of these planets are thought to contain a variety of species
 not considered previously \cite{15ItIkKa.exo,16FeJaWi,jt693,20HeWoHe}. 
 Secondly higher  temperature
 planets have been observed than anyone anticipated, the poster child for hot exoplanets
 is Kelt-9b which is thought to reach temperatures of 4000 K \cite{17GaStCo.exo},
 hotter than most of the stars in our Galaxy! This has led to the need to construct
 line list for key species over an extended temperature range.
 Third is the development of
 Doppler-shift high resolution spectroscopy of exoplanets \cite{14Snellen,18Birkby} which has proved
 a powerful tool for detecting molecules but fails in the absence of highly
 accurate molecular line lists \cite{15HoDeSn.TiO}. These developments have led us
 to both expand the range of molecules included in the database and to begin
 a systematic attempt to improve the accuracy of the line positions for the line
 lists contained in the database. Progress on both of these objectives is described
 below as well work on extending the coverage of the database into the ultraviolet.

 Of course ExoMol is not the only database providing spectroscopic data for
 atmospheric studies. For the Earth's atmosphere, databases HITRAN \cite{jt350,jt453,jt546} and GEISA
\cite{jt504,jt636} provide comprehensive and validated datasets for approximately
50 key molecules. However, these databases are designed for studies at room temperature and below and do not
contain the necessary data for adequately calculating radiative transport in hot bodies. The HITEMP
database was constructed to extend HITRAN to higher temperatures. The latest,
2010, release \cite{jt480} only contained data on 5 molecules (OH, NO, CO, H$_2$O and CO$_2$) 
and improved hot line lists are available for all these molecules; an update to HITEMP is
currently in progress \cite{jt763,20HaGoRe.CH4,jtH2OHITEMP2020}. 
Other relevant databases include TheoReTS \cite{TheoReTS}, which contains hot line lists for 8 polyatomic molecules,
and  Kurucz's compilation of data, which is very complete for atomic sources but contains data only on about ten diatomic molecules 
\cite{11Kurucz.db} all of which are covered by ExoMol, usually to higher accuracy. The MoLLIST data base of Bernath and co-workers
\cite{MOLLIST} contains empirically derived line lists for about 20 diatomic species. These line lists
have recently been incorporated into the ExoMol database  \cite{jt790}, see Table~\ref{tab:otherdata} below. Databases
of hot molecular spectra for other specialised applications include
those for combustion \cite{12RiSoxx,15CoLixx}
and studies of laser-induced plasmas
\cite{15PaWoSu}.
ExoMol line lists together with data from other sources will serve as the main data source for exoplanet stands
planned for the upcoming space missions Ariel and JWST.

The methodology \cite{jt511,jt573,jt624,jt632,jt654,jt656,jt693} and software \cite{jt609,jt626} developed and used by the ExoMol project has been
extensively documented elsewhere. Here we will only consider those aspects
which impinge on the new ExoMol2020 release.

\section{Database coverage}

The ExoMol project aims at complete coverage of the spectroscopic
properties of molecules which are deemed to be important in hot
astrophysical environments. Coverage concerns (a) the molecular
species considered, including isotopologues; (b) the spectroscopic and thermodynamic properties 
considered; (c) the frequency range
considered and (d) the upper temperature range for which the data are
reasonably complete. Both the required temperature and frequency range
completeness are to some extent a judgement on what is required for
astronomical and other studies.

 Unlike ExoMol2016, the ExoMol2020
database essentially provides a complete dataset for modelling hot atmospheres.
While new species are still being added, commonly at the request of users,
the database now contains molecular data for  80 molecules and covers the key molecules thought to be important
for exoplanetary studies. In many cases
there is more than one line list for a given isotopologue. For this
reason it is our practice to name each line list, including (unnamed) ones
we have imported from other sources. In the case of multiple line lists the ExoMol
website provides a recommended line list for the given species. This recommendation
is for studies of hot atmospheres; other available line lists
may be more suitable for room temperature studies.

For ease of understanding
we have divided the description of line lists provided into three tables.
Tables \ref{tab:exomoldata} and \ref{tab:otherdata} summarise the molecules
for which data are provided by ExoMol and give the characteristics of the
line list in each case. 
Table \ref{tab:exomoldata} details  line lists which have been
formally published as part of the ExoMol project while Table~\ref{tab:otherdata}
shows line lists imported  from other sources which have been recast in the ExoMol 
format (see Wang {\it et al} \cite{jt790} for example); these line lists
are fully integrated into the database.
Table  \ref{tab:otherdata2} lists the main
extra line lists available through the ExoMol website; this table is not comprehensive but
the other line lists available on the ExoMol website and not listed in Table 1 -- 3 are largely of historic interest only.

\begin{table}[H]
\tiny

\caption{Datasets created by the ExoMol project and included in the ExoMol
database.}
\label{tab:exomoldata}
\begin{tabular}{llrrrrlcl}
\hline\hline
Paper&Molecule&$N_{\rm iso}$&$T_{\rm max}$&$N_{\rm elec}$&$N_{\rm lines}$$^a$&DSName&&Reference\\
\hline
I&BeH&1&2000 &1&16~400&Yadin& &\citet{jt529}\\
I&MgH&3 &2000 &1&10~354&Yadin&\tick&\citet{jt529}\\
I&CaH&1 &2000 &1&15~278&Yadin&\tick&\citet{jt529}\\
II&SiO&5&9000&1& 254~675&EJBT&\tick&\citet{jt563}\\
III&HCN/HNC&1$^a$&4000&1&399~000~000&Harris&\tick&\citet{jt570}\\
IV&CH$_4$&1&2000&1&34~153~806~005&YT34to10&\tick&  \citet{jt564,jt698}\\
V&NaCl&2&3000&1& 702~271 &Barton&\tick&\citet{jt583}\\
V&KCl&4&3000&1& 1~326~765  &Barton&\tick&\citet{jt583}\\
VI&PN&2&5000&1&142~512&YYLT&\tick&\citet{jt590}\\
VII&PH$_3$&1&1500&1&16~803~703~395&SAlTY&\tick&\citet{jt592}\\
VIII&H$_2$CO&1&1500&1&10~000~000~000&AYTY&\tick&\citet{jt597}\\
IX&AlO&4&8000&3&4~945~580&ATP&\tick&\citet{jt598}\\
X&NaH&2&7000&2&79~898&Rivlin&\tick&\citet{jt605}\\
XI&HNO$_3$&1&500&1&6~722~136~109&AlJS&\tick&\citet{jt614}\\
XII&CS&8&3000&1&548~312&JnK&\tick&\citet{jt615}\\
XIII&CaO&1&5000&5&21~279~299&VBATHY&\tick&\citet{jt618}\\
XIV&SO$_2$&1&2000&1&1~300~000~000&ExoAmes&\tick&\citet{jt635}\\
XV&H$_2$O$_2$&1&1250&1&20~000~000~000&APTY&\tick&\citet{jt638}\\
XVI&H$_2$S&1&2000&1&115~530~3730& AYT2&\tick&\citet{jt640}\\
XVII&SO$_3$&1&800&1&21~000~000~000&UYT2&\tick&\citet{jt641}\\
XVIII&VO&1&2000&13&277~131~624&VOMYT&\tick&\citet{jt644}\\
XIX&H$_2$$^{17,18}$O&2&3000&1&519~461~789&HotWat78&\tick&\citet{jt665}\\
XX&H$_3^+$&1$^b$&3000&1&11~500~000~000&MiZATeP&\tick&\citet{jt666}\\
XXI&NO&6&5000&2&2~281~042&NOName&\tick&\citet{jt686}\\
XXII&SiH$_4$&1&1200&1&62~690~449~078&OY2T&\tick&\citet{jt701}\\
XXIII&PO&1&5000&1&2~096~289&POPS&\tick&\citet{jt703}\\
XXIII&PS&1&5000&3&30~394~544&POPS&\tick&\citet{jt703}\\
XXIV&SiH&4&5000&3&1~724~841&SiGHTLY&\tick&\citet{jt711}\\
XXV&SiS&12&5000&1&91~715&UCTY&\tick&\citet{jt724}\\
XXVI&NS&6&5000&1&3~479~067&SNaSH&\tick&\citet{jt725}\\
XXVI&HS&6&5000&1&219~463&SNaSH& &\citet{jt725}\\
XXVII&C$_2$H$_4$&1&700&1&60~000~000~000&MaYTY&\tick&\citet{jt729}\\
XXVIII&AlH&3&5000&3&40~000&AlHambra&\tick&\citet{jt732}\\
XXIX&CH$_3$Cl&2&1200&1&166~279~593~333& OYT &\tick&\citet{jt733} \\
XXX&H$_2$$^{16}$O&1$^c$&5000&1&5~745~071~340&POKAZATEL&\tick&\citet{jt734}\\
XXXI&C$_2$&3&5000&8& 6~080~920&8states&\tick&\citet{jt736}\\
XXXII&MgO&5&5000&5&72~833~173&LiTY&\tick&\citet{jt759}\\
XXXIII&TiO&5&5000&13&59~000~000&Toto&\tick&\citet{jt760}\\
XXXIV&PH&1&4000&2&65~055&LaTY&\tick&\citet{jt765}\\
XXXV&NH$_3$&1$^d$&1500&1&16~900~000&CoYuTe&\tick&\citet{jt771}\\
XXXVI&SH&2&3000& 2&572~145&GYT&\tick&\citet{jt776}\\
XXXVII&HCCH&1&2000&1&4~347~381~911&aCeTY &\tick&\citet{jt780}\\
XXXVIII&\ce{SiO2}&1&3000&1&32~951~275~437&OYT3 &\tick&\citet{jt797}\\
XXXIX&\ce{CO2}&1& 3000 &1  & 7~996~570~390  &UCL-4000 &\tick&\citet{jt804}\\
XL&\ce{H3O+}&1&1500&1 & 2~089~331~073 &eXeL &\tick&\citet{jt805}\\
\hline
\hline
\end{tabular}

Paper Number in series published in Mon. Not. R. Astron. Soc.\\
$N_{\rm iso}$ Number of isotopologues considered;\\
$T_{\rm max}$ Maximum temperature for which the line list is complete;\\
$N_{\rm elec}$ Number of electronic states considered;\\
$N_{\rm lines}$  Number of lines: value is for the main isotope.\\
\tick\ indicates line list recommended  for studies of hot atmospheres.\\
$^a$ The Larner line list for H$^{13}$CN/HN$^{13}$C due to \citet{jt447} is recommended.\\
$^b$ The ST line list for H$_2$D$^+$ due to
\citet{jt478} is recommended,.\\
$^c$ The VTT line list for HDO due to \citet{jt469} is recommended.\\
$^d$ There is a room temperature $^{15}$NH$_3$ line list due to 
\citet{15Yurche.NH3}.\\
\end{table}

\begin{table}[H]
\caption{Datasets not created as part of the ExoMol project but included in the
ExoMol database.}
\label{tab:otherdata}
\tiny
\begin{tabular}{lcrcrlcll}
\hline\hline
Molecule&$N_{\rm iso}$&$T_{\rm max}$&$N_{\rm elec}$&$N_{\rm lines}$&DSName&&Reference&Methodology\\
\hline
\ce{H2}	&	1	&10000&	1	&	4712	&RACPPK	&\tick&\citet{19RoAbCz.H2}& Ab initio \\
CH	&	1	&5000&	4	&	52201	&	MoLLIST&\tick&\citet{14MaPlVa.CH}&Empirical \\
NH	&1&	 5000     &1&		12150	&	MoLLIST&\tick&\citet{14BrBeWe.NH,15BrBeWe.NH,18FeBeHo.NH} &Empirical \\
OH  & 1 & 5000 & 2 & 54276 & MoLLIST &\tick&\citet{16BrBeWe.OH} &Empirical\\
AlCl	&	2	& 5000& 1	&	20245	&		MoLLIST&\tick&\citet{18YoBexx.AlF} &Empirical\\
AlF	&		1	&5000	& 1	&	40490	&	MoLLIST&\tick&\citet{18YoBexx.AlF}&Empirical \\
OH$^+$	&	1	&5000	& 2	&		12044	&	MoLLIST	&\tick&\citet{17HoBexx.OH+,18HoBiBe.OH+}&Empirical \\
CaF	&	1&5000	& 6		&	14817	&	MoLLIST&\tick&\citet{18HoBexx.CaF}&Empirical \\
MgF	&1	&	5000& 3	&	8136	&	MoLLIST&\tick&\citet{17HoBexx.MgF}&Empirical \\
KF	&1 &	5000&	2	&	10572	&		MoLLIST&\tick&\citet{16FrBeBr.NaF}&Empirical \\
NaF	&	1 &5000 & 	1	&	7884	&		MoLLIST&\tick&\citet{16FrBeBr.NaF}&Empirical \\
LiCl	&	1&	5000& 4	&	26260	&		MoLLIST&\tick&\citet{18BiBexx.LiF}&Empirical \\
LiF	&	1	& 5000&	2	&	10621	&	MoLLIST&\tick&\citet{18BiBexx.LiF}&Empirical \\
MgH	&	2	&5000	&1	&	14179	&	MoLLIST&\tick&\citet{13GhShBe.MgH}&Empirical \\
TiH	&	1 &	5000&	3	&	181080	&	MoLLIST&\tick&\citet{05BuDuBa.TiH}&Empirical \\
CrH	&	1&	5000&2	&	13824	&		MoLLIST&\tick&\citet{06ChMeRi.CrH} &Empirical\\
FeH	&	1 &5000	&	2		&	93040	&	MoLLIST&\tick&\citet{10WEReSe.FeH} &Empirical\\
HF	&	2& 5000& 1 &	7956	&  Coxon-Hajig	&\tick&\citet{15CoHaxx.HF} &Empirical\\
HCl	&	4& 5000& 1 &	2588	&	HITRAN&\tick&\citet{11LiGoBe.HCl} &Empirical\\
CP	&	1& 5000&	2&	28752	&		MoLLIST&\tick&\citet{14RaBrWe.CP} &Empirical\\
CN	&	1 &	5000&	3	&	195120	&		MoLLIST& \tick&\citet{14BrRaWe.CN} &Empirical\\
C$_2$&1		&	5000&2	&	47~570	&	MoLLIST& &\citet{13BrBeSc.C2}&Empirical \\
CaH	&	2& 5000&1&	6000	&	MoLLIST&\tick&\citet{12LiHaRa.CaH,13ShRaBe.CaH} &Empirical\\
N$_2$	&	1&10000 & 4$^a$&	7~182~000	&	WCCRMT	&\tick&\citet{18WeCaCr.N2} &Empirical\\
SiO&1&5000&3&6~67~814&Kurucz-SiO&&\citet{11Kurucz.db}&Empirical\\
ScH&1&5000&6&1~152~827&LYT&\tick&\citet{jt599}&Ab initio\\
LiH&1&12000&1&18~982&CLT&\tick&\citet{jt506}&Ab initio\\
LiH$^+$&1&12000&1&332&CLT&\tick&\citet{jt506}&Ab initio\\
CO&9&9000&1&752~976&Li2015&\tick&\citet{15LiGoRo.CO}&Empirical\\
HeH$^+$&4&9000&1&1430&ADSJAAM&\tick&\citet{19AmDiJo}&Ab initio\\
HD$^+$&1&9000&1&10285&ADSJAAM&\tick&\citet{19AmDiJo}&Ab initio\\
HD&1&9000&1&5939&ADSJAAM&\tick&\citet{19AmDiJo}&Ab initio\\
CH$_3$F&1&300&1&139~188~215&OYKYT&\tick&\citet{19OwYaKu.CH3F}&Ab initio\\
AsH$_3$&1&300&1&3~600~000&CYT18&\tick&\citet{jt751}&Ab initio\\
P$_2$H$_2$$^b$&2&300&1&10~667~208~951&OY-Trans&\tick&\citet{19OwYuxx.P2H2}&Ab initio\\
P$_2$H$_2$$^b$&2&300&1&11~020~092~365&OY-Cis&\tick&\citet{19OwYuxx.P2H2}&Ab initio\\
PF$_3$&1&300&1&68~000~000~000&MCYTY&\tick&\citet{jt752}&Ab initio\\
CH$_3$&1&1500&1&2~058~655~166&AYYJ&\tick&\citet{19AdYaYu.CH3}&Ab initio\\
BeH&3&5000 &2&592308&Darby-Lewis&\tick&\citet{jt722}&ExoMol\\
CO$_2$&13&4000&1&298~323~789&Ames-2016& &\citet{13HuFrTa.CO2,Huang2017}&ExoMol-like\\
SiH$_2$&1&2000&1&254~061~207&CATS &\tick&\citet{jt779} &ExoMol\\
YO&1&5000&6&3520133&SSYT&\tick&\citet{jt774}&Ab initio\\
\hline\hline
\end{tabular}

$N_{\rm iso}$ Number of isotopologues considered;\\
$T_{\rm max}$ Maximum temperature for which the line list is complete;\\
$N_{\rm elec}$ Number of electronic states considered;\\
$N_{\rm lines}$  Number of lines: value is for the main isotope.\\
\tick\ indicates line list recommended for studies of hot atmospheres.\\
$^a$ The WCCRMT line list considers triplet states only.\\
$^b$ There are separate line lists for cis and trans P$_2$H$_2$.\\
\end{table}

\begin{table}[H]
\caption{Supplementary datasets available from the website.}
\label{tab:otherdata2}
\tiny
\begin{tabular}{lcrcrlcll}
\hline\hline
Molecule&$N_{\rm iso}$&$T_{\rm max}$&$N_{\rm elec}$&$N_{\rm lines}$&DSName&&Reference&Methodology\\
\hline
H$_3^+$&2$^a$&4000&1&3~070~571&NMT& &\citet{jt181}&ExoMol\\
H$_2$O &2$^b$&3000&1&505~806~202&BT2& &\citet{jt378}&ExoMol\\
NH$_3$&2$^c$&1500&1&1~138~323~351&BYTe& &\citet{jt500}&ExoMol\\
HeH$^+$&4&10000&1&1~431&Engel& &\citet{jt347}&Ab initio\\
HD$^+$&1&12000&1&10~119&CLT& &\citet{jt506}&Ab initio\\
CO$_2$&13&300 & 1&  161944 & Zak & &\citet{jt625,jt667,jt678}&ExoMol\\
CO$_2$&1&4000&1&628~324~454&CSSD-4000& &\citet{11TaPexx.CO2}&Empirical\\
H$_2$O&1&300&1& & WAT\_UV296 & \cite{jt803}&ExoMol\\
\hline
\hline
\end{tabular}

$N_{\rm iso}$ Number of isotopologues considered;\\
$T_{\rm max}$ Maximum temperature for which the line list is complete;\\
$N_{\rm elec}$ Number of electronic states considered;\\
$N_{\rm lines}$  Number of lines: value is for the main isotope.\\

\end{table}

\section{Individual line lists}

\begin{figure}[t]
\begin{center}
\includegraphics[width=0.75\textwidth]{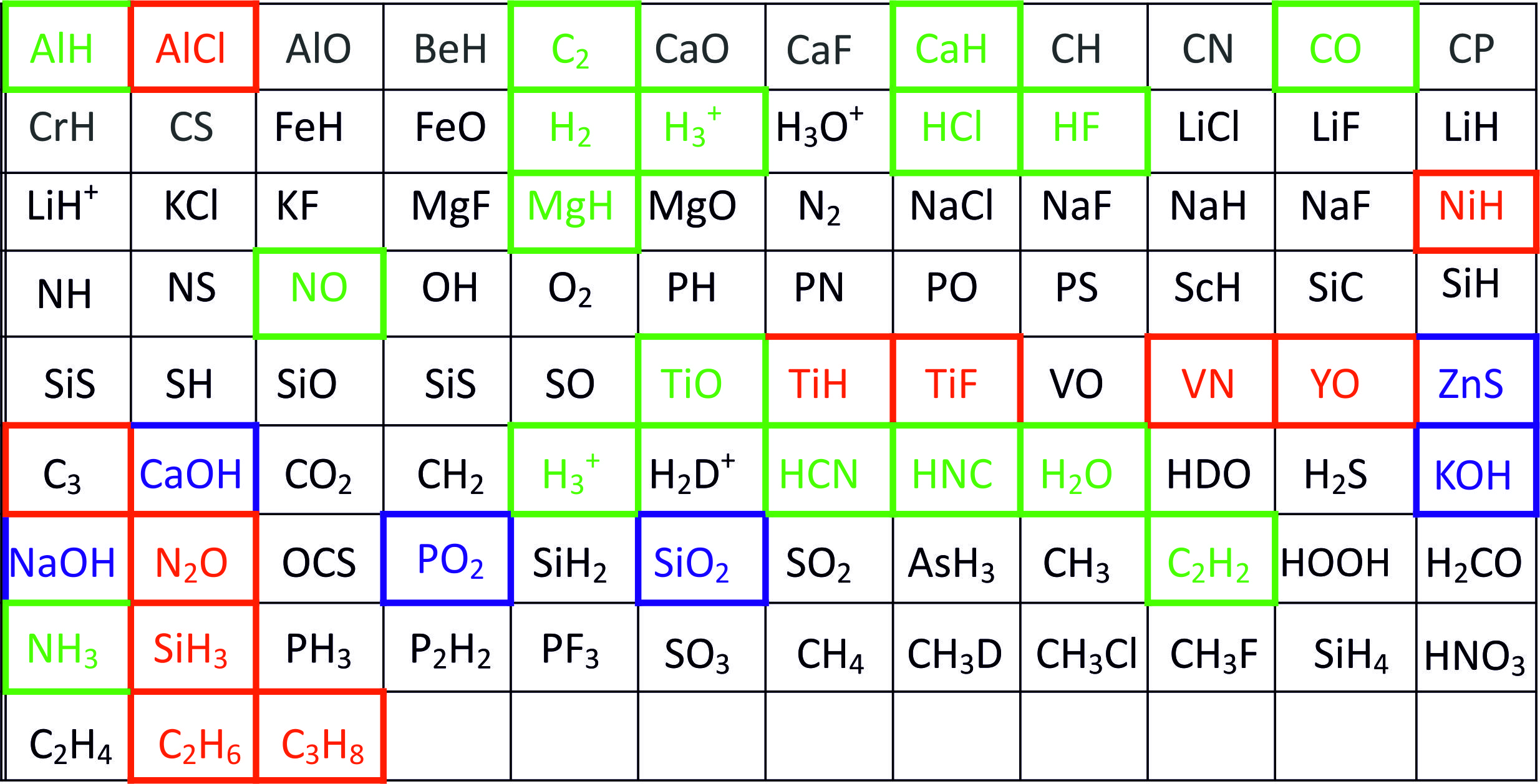}
\end{center}
\caption{ Molecular line lists: red indicates line lists in progress, blue corresponds to the line lists suggested for molecules specific for hot rocky exoplanets and green indicates line lists which contain data applicable for high resolution.}
\label{fig:linelist}
\end{figure}

An overview of the line lists in the ExoMol database is given in Fig.~\ref{fig:linelist}.

One general issue is that Medvedev and co-workers
\cite{highv,15LiGoRo.CO} identified a numerical problem with the
intensities of high overtone transitions computed with the
standard compilation of the diatomic vibration-rotation program {\sc Level}
\cite{LEVEL}.  Our diatomic line lists computed with {\sc Level} or {\sc Duo} \cite{jt608} have been adjusted
to remove transitions affected by this issue. Medvedev {\it et al.} \cite{jt794} recently identified similar
issues with triatomic systems but tests suggest that in practice our triatomic line lists seem largely
unaffected by the problem.

Below we consider some of the line lists presented in the ExoMol database and listed in Tables \ref{tab:exomoldata} and \ref{tab:otherdata}. We restrict our discussion to issues not
covered in the ExoMol2016 release or the original publications. We start by considering the 42 molecules for which line lists have been created as part of the ExoMol series
as listed in Table \ref{tab:exomoldata}.

\subsection{Diatomics}

\subsubsection{AlH, paper XVIII}

There is a new AlHambra line list for AlH \cite{jt732}. A MARVEL (measured active rotational-vibrational energy levels
\cite{jt412}) project was performed as part of this project meaning that many transitions are predicted with experimental
accuracy. The  AlHambra line list has been updated to give the uncertainty in the energy in the states file which
allows users to determine the uncertainty in a given transition wavenumber.

\subsubsection{AlO, paper IX}

No change. The APTY line list was recently used to make a detection of AlO in an exoplanet atmosphere, 
the ultra-hot Jupiter exoplanet WASP-33~b \cite{19VaMaWe.AlO} and hot Jupiter exoplanet WASP-43~b \citep{20ChMiKa}.

\subsubsection{BeH, paper I}

The BeH \lq\lq Yadin\rq\rq\ line list was one of the first  constructed by ExoMol \cite{jt529}; however,
the line list only included transitions within the
X~$^2\Sigma^+$ ground electronic state. Recently 
\citet{jt722} constructed line lists for BeH, BeD and BeT which consider both
the ground and first excited (A~$^2\Pi$) states. At the same time Darby-Lewis
{\it et al.} performed a MARVEL analysis  and used their empirical levels in the fit. 
The Darby-Lewis line list is therefore more accurate and more extensive than the Yadin one which it
replaces in the ExoMol database and  is therefore recommended. Yadin is still available on the website though renamed to Yadin-BeH in order to avoid conflict with the ``Yadin'' line lists for MgH and CaH, which remain to be recommended for the IR region.

\subsubsection{C$_2$, paper XXXI}

There are two new line lists for C$_2$ in the ExoMol2020 release: ExoMol 8state \cite{jt736} and the empirical MoLLIST \cite{13BrBeSc.C2}. The MoLLIST line list
only considers the much-observed Swan band while 8state covers the 8 band systems
which interconnect the 8 lowest electronic states of C$_2$. As 8state uses empirical
MARVEL energy levels \cite{jt637} where available it should be as accurate as MoLLIST
for the Swan band. Use of 8state is therefore recommended. Since MoLLIST line lists  are  recommended for other molecules and in order to avoid conflicts with the `recommended' flag for 8states, the C$_2$ MoLLIST line list is now referenced to as MoLLIST-C2 on ExoMol. 

An update of the C$_2$ MARVEL data has just been completed \cite{jt809}; this has been used to improve
the 8state energies and hence transition wavenumbers. This latest version includes uncertainties in the states file.

\subsubsection{CaH, Paper I}
The \lq\lq Yadin\rq\rq\ CaH line list only considers transitions within the
X~$^2\Sigma^+$ ground electronic state \cite{jt529}. MoLLIST provides 
a rovibronic line list for the $X\,^2\Sigma^+$--$X\,^2\Sigma^+$, 
$A\,^2\Pi$--$X\,^2\Sigma^+$, $B\,^2\Sigma^+$--$X\,^2\Sigma^+$ and E~$^2\Pi$ -- X~$^2\Sigma^+$  systems  due to
\citet{11RaTeGo.CaH,12LiHaRa.CaH,13ShRaBe.CaH}. 
Both line lists are included in the ExoMol database; work is in progress on creating
a single unified line list which will extend the range of rovibronic transitions considered. In the meantime both line lists
are recommended.

\subsubsection{CaO, Paper XIII}

No change.

\subsubsection{CS, Paper XII}

No change. We note that a line list for CS has also recently been supplied by \citet{20HoWexx.CS} and
\citet{20XiShSu.CS} have extended consideration to rovibronic transitions for the lowest, A~$^1\Pi$ -- X~$^1\Sigma^+$,
allowed electronic band.

\subsubsection{KCl, Paper V}

No change

\subsubsection{MgH, Paper I}
The ExoMol \lq\lq Yadin\rq\rq\  line list for MgH only considers transitions within the
X~$^2\Sigma^+$ ground electronic state \cite{jt529}. MoLLIST provides 
a rovibronic line list containing A~$^2\Pi$ -- X~$^2\Sigma^+$ and B$^\prime$~$^2\Sigma^+$
-- X~$^2\Sigma^+$
transitions due to \citet{13GhShBe.MgH}.
Both line lists are included in the ExoMol database; work is in progress on creating
a single unified line list which will extend the range of rovibronic transitions considered.
In the meantime both line lists
are recommended.

\subsubsection{MgO, Paper XXXII}

There is a new LiTY line list for MgO \cite{jt759}.

\subsubsection{NaCl, Paper V}

No change.

\subsubsection{NaH, Paper X}

No change.

\subsubsection{NO, XXI}

There is a new NOName line list for NO \cite{jt686}. This line list was constructed
using a combination of standard ExoMol and empirical methodologies, and also included a MARVEL project. Its transition frequencies should therefore be close to experimental accuracy.
NOName has largely been adopted for the new release of HITEMP \cite{jt763}.

NOName only covers transitions between levels which lie within the X~$^2\Pi$ ground electronic state of NO. A new line list which includes ultraviolet rovibronic transitions of NO is currently being constructed.

\subsubsection{PH, Paper XXXIV}

There is a new LaTY line list for PH \cite{jt765}.

\subsubsection{PN, Paper VI}

No change. An extended line list covering visible and ultra violet (UV), accompanied by a MARVEL project is currently in progress.

\subsubsection{PO, Paper XXIII}

There is a new POPS line list for PO \cite{jt703}.

\subsubsection{PS, Paper XXIII}

There is a new POPS line list for PS \cite{jt703}.

\subsubsection{SH, Paper XXXVI}

There are two new ExoMol line lists for SH: SNaSH  \cite{jt725} which
only covers transitions within the X~$^2\Pi$ ground
electronic state of SH and the newer GYT line list of \citet{jt776}.
GYT considers both transitions within  X~$^2\Pi$ and vibronic transitions in the A~$^2\Sigma^+$ -- X~$^2\Pi$ band system. As GYT also improves the accuracy
of the X state transitions, it is recommended for all applications. 
The SNaSH SH line list  is also renamed to SNaSH-SH in order to avoid a conflict  and retained only for completeness. 

\subsubsection{SiO, Paper II}

The ExoMol SiO line lists only consider transitions within the
X~$^1\Sigma^+$ ground electronic state  \cite{jt563}; these line lists
have been widely used including a recent determination of isotopologue ratios in Arcturus \cite{jt777}. For this release of ExoMol, an empirical and less accurate SiO line list by \citet{11Kurucz.db} covering the X--X, A--X and E--X electronic bands was added. 
Given the importance of
SiO for lava planets \cite{jt693}, construction of a comprehensive (accurate and complete) rovibronic line list for SiO covering both IR and ultraviolet would be useful.

\subsubsection{SiH, Paper XXIV}

There is a new SiGHTLY line list for SiH \cite{jt711}.

\subsubsection{SiS, Paper XXV}

There is a new UCTY line list for SiS \cite{jt724}.

\subsubsection{SN, Paper XXVI}

There is a new SNaSH line list for SN \cite{jt725}.

\subsubsection{TiO, Paper XXXIII}

There is a new Toto line list for TiO \cite{jt760}. The line list for the main isotopologue ($^{48}$Ti$^{16}$O) used empirical energies from
an associated MARVEL study \cite{jt672} while the energy levels of other isotopologues were improved
using the procedure of \citet{jt665}. The Toto line list has been updated to give MARVEL uncertainties, where available, in the
States file. For other states the uncertainties were estimated as follows:  1 \cm\ for all levels in  vibronic states that contain levels determined by MARVEL, 10 \cm\ for levels in those electronic states who excitation energy ($T_e/T_0$) is  known from experiment and 100 \cm\ for all other levels. 

TiO is a particularly important species in the atmosphere of 
cool stars and has been detected in exoplanets \cite{17SeBoMa.TiO,17NuKaHa.TiO}. The Toto line list is
significantly better at reproducing stellar spectra than the line lists due to 
\citet{98Plxxxx.TiO} and \citet{98Scxxxx.TiO} which it supersedes. Furthermore a recent
study of Ti isotope abundances in two M-dwarf stars demonstrated the accuracy of the line lists
for several  isotopologues of TiO \cite{jt799}. However, TiO remains a challenging 
system to treat using {\it ab initio} electronic structure methods (see \citet{jt623}) and further work is
required before there is a definitive line list for TiO.

\subsubsection{VO, Paper XVIII}

There is a new VOMYT line list for VO \cite{jt644} which has been used to tentatively identify VO in the
atmosphere of exoplanets \cite{16EvSiWa.VO,jt699}. We note that VO spectra display particularly large splittings
due to hyperfine effects as well as transitions which are only allowed due to these effects \cite{82ChHaMe.VO}.
VOMYT does not include hyperfine effects in its spectroscopic model meaning that the line list unsuitable for
high resolution studies. We plan to address this problem by both performing a hyperfine resolved MARVEL study
and extending {\sc Duo} to allow for hyperfine effects.

\subsection{Triatomics}

\subsubsection{CO$_2$, Paper XXXIX}
Line lists for hot CO$_2$ have been constructed by NASA Ames \cite{13HuFrTa.CO2,13HuFrTa.CO2} using methodologies similar
to those employed by ExoMol. 
The CDSD-4000 hot line list due to \citet{11TaPexx.CO2} and the room temperature line lists due to 
 \citet{jt625,jt667,jt678} are also available on the ExoMol website. The
CDSD-4000 line list is  based on the use of an effective Hamiltonian which tends to  miss contributions from  unobserved hot bands. 
 
A new ExoMol line list UCL-4000 for \ce{CO2} has been produced \cite{jt804} using an accurate \textit{ab initio} dipole moment surface (DMS) by \citet{jt613} and empirical potential energy surface (PES) Ames-2016 by \citet{17HuScFr.CO2}; where possible computed energies have been replaced by empirical one
derived from HITRAN. The UCL-4000 line list covers the wavenumber range 0--20000~\cm\ and should be applicable for temperatures up to 4000~K \citep{jt804}.  It is recommended for use in high temperature applications.

\subsubsection{HCN/HNC, Paper III}

The combined HCN/HNC ExoMol line list of Barber \ea\ \cite{jt570}
made extensive use of empirical energy levels due to Mellau \cite{11Mexxxx.HCN,
11Mexxxx.HNC}. This has enabled the line list to be successfully used in
high-resolution Doppler spectroscopy studies of exoplanets \cite{18HaMaCa.HCN,19CaMaHa.HCN,jt782}.

\subsubsection{H$_2$O, Papers XIX and XXX}

Water is one of the most widely studied molecules and its spectrum
has been detected in a variety of exoplanets, in many cases as the only
clearly identifiable molecule \cite{jt699,jt739}. Good water line lists
have been available for some time, in particular the Ames line list
 of Partridge
and Schwenke \cite{97PaScxx.H2O} and the BT2 line list due to \citet{jt378}.
While BT2 was significantly more complete than Ames, the Ames line list
was more accurate, particularly at infrared wavelengths. A new line list,
known as POKAZATEL \cite{jt734} has been generated by the ExoMol
project. POKAZATEL is complete in the sense that it contains transitions between
all bound rotation-vibration states in the molecule and thus is reliable for temperatures above 3000 K where
the earlier line lists are not valid. POKAZATEL is also intrinsically more accurate
than the Ames line list and, indeed, as many of its energy levels have been
replaced by empirical energy levels generated using MARVEL \cite{jt539}, key
transition frequencies are actually reproduced to experimental accuracy.
POKAZATEL should therefore be used in preference to the earlier line lists.

It has become apparent that the rotation-vibration spectrum of water is responsible for weak but observable near-UV
absorption in the Earth's atmosphere \cite{jt645}. The POKAZATEL line list, which used the {\it ab initio} LTP2001S DMS of \citet{jt509}, appears to underestimate the strength of water absorption in the blue and near-UV. As a response to this
 Conway and co-workers have recently developed a  more accurate global dipole moment surface \cite{jt744}. Through a significant number of comparisons against high quality experimental and theoretical sources of spectroscopic data \cite{jt775,jt785}, their results suggest there may be advantages of using their line lists in particular regions, primarily at short wavelengths. For this reason, a new room temperature H$_2$$^{16}$O line list called WAT\_UV296 \cite{jt803} has been computed. 
 This line list is available on the website and should be used for
 room temperature studies of spectra with $\lambda < 0.5$~$\mathrm{\mu m}$. A high temperature companion to this line list will form the basis of the updated HITEMP database \cite{jtH2OHITEMP2020}. 
 

\citet{jt665} provided the HotWat78 line lists for H$_2$$^{17}$O and H$_2$$^{18}$O.
The energy levels of this line list were  improved using MARVEL energies
\cite{jt454,jt482}. Furthermore, \citet{jt665} developed a method which gives
excellent isotopologue energy levels for states only observed for  H$_2$$^{16}$O.
Finally the VTT line list due to \citet{jt469} remains the recommended one for HDO; a new, improved HDO line list is currently
under construction \cite{jtHDO}.

A new room temperature line list CKYKKY  for \ce{H2O} containing electric quadrupole moments was computed using an \textit{ab initio} quadrupole moment surface \citep{20CaKaYa.H2O} and accurate empirical PES by \citet{jt714} with the variational program TROVE \cite{TROVE}. The energies were replaced by the \ce{H2O} IUPAC MARVEL values \citep{jt539} or HITRAN values \citep{jt691}, if available.

\subsubsection{H$_2$S, Paper XVI}

There is a new ExoMol AYT2 line list for hydrogen sulphide due to \citet{jt640}.
Since completion of this line list a MARVEL project has been performed for
H$_2$S \cite{jt718}; the AYT2 line list is being updated to use these
improved energy levels.

\subsubsection{H$_3^+$}

The new ExoMol MiZATeP line list of \citet{jt666} replaces that of
 \citet{jt181} (NMT). The new line list uses empirical energy levels
 from the MARVEL study of \citet{13FuSzFa.H3+}. Astronomy  makes wide use of H$_3^+$ line lists and is reliant on {\it ab initio} line intensities since
 no absolute line intensities have been measured in the laboratory, see \citet{jt587}. This astronomical work
 on H$_3^+$ has recently been reviewed
 by \citet{jt800}. The ST \ce{H2D+} line list due to \citet{jt478} plus the newly generated \ce{D3+} and \ce{D2H+} line lists
 will be MARVELised and released in the near future.
 
\subsubsection{SO$_2$, Paper XVII}

The ExoAmes line list for SO$_2$ \cite{jt637} is unchanged since the ExoMol2016 release. However, a 
MARVEL set of energy levels for SO$_2$ are now available \cite{jt704}
and will be used to update
ExoAmes in the near future.

\subsection{Tetratomics}

\subsubsection{HCCH, Paper XXXVII}

The new aCeTY line list for acetylene due to \citet{jt780} has recently been released. This line list already incorporated the
MARVEL energies of \citet{jt705} and gives uncertainties in the energy levels as part of the States file.

\subsubsection{H$_2$CO, Paper VIII}

The AYTY formaldehyde line list \cite{jt597} is in the process of being updated with empirical energy levels produced by 
a recent MARVEL study \cite{jtH2COmarvel};  this will make the line list suitable for high resolution studies.

\subsubsection{H$_2$O$_2$, Paper XV}

There is a new ExoMol APTY line list for hydrogen peroxide \citep{jt620}.
We note because of difficulties with assigning spectra, even at room temperature
HITRAN is very incomplete for 
H$_2$O$_2$. APTY should give complete coverage at infrared wavelengths.
Recently APTY was used to suggest the importance of \ce{H2O2} as a greenhouse gas on oxidised Early Mars \citep{20ItHaTa.H2O2}.

\subsubsection{H$_3O^+$, Paper XL}
A line list for the hydronium ion, H$_3$O$^+$, has recently been constructed \cite{jt805} in response to a laboratory study by
\citet{jt793} which suggested both that hydronium is likely to be a dominant molecular ion in gaseous
exoplanets and that it should be detectable by upcoming space missions.

\subsubsection{NH$_3$, Paper XXXV}

\begin{figure}[H]
\begin{center}
\includegraphics[width=0.90\textwidth]{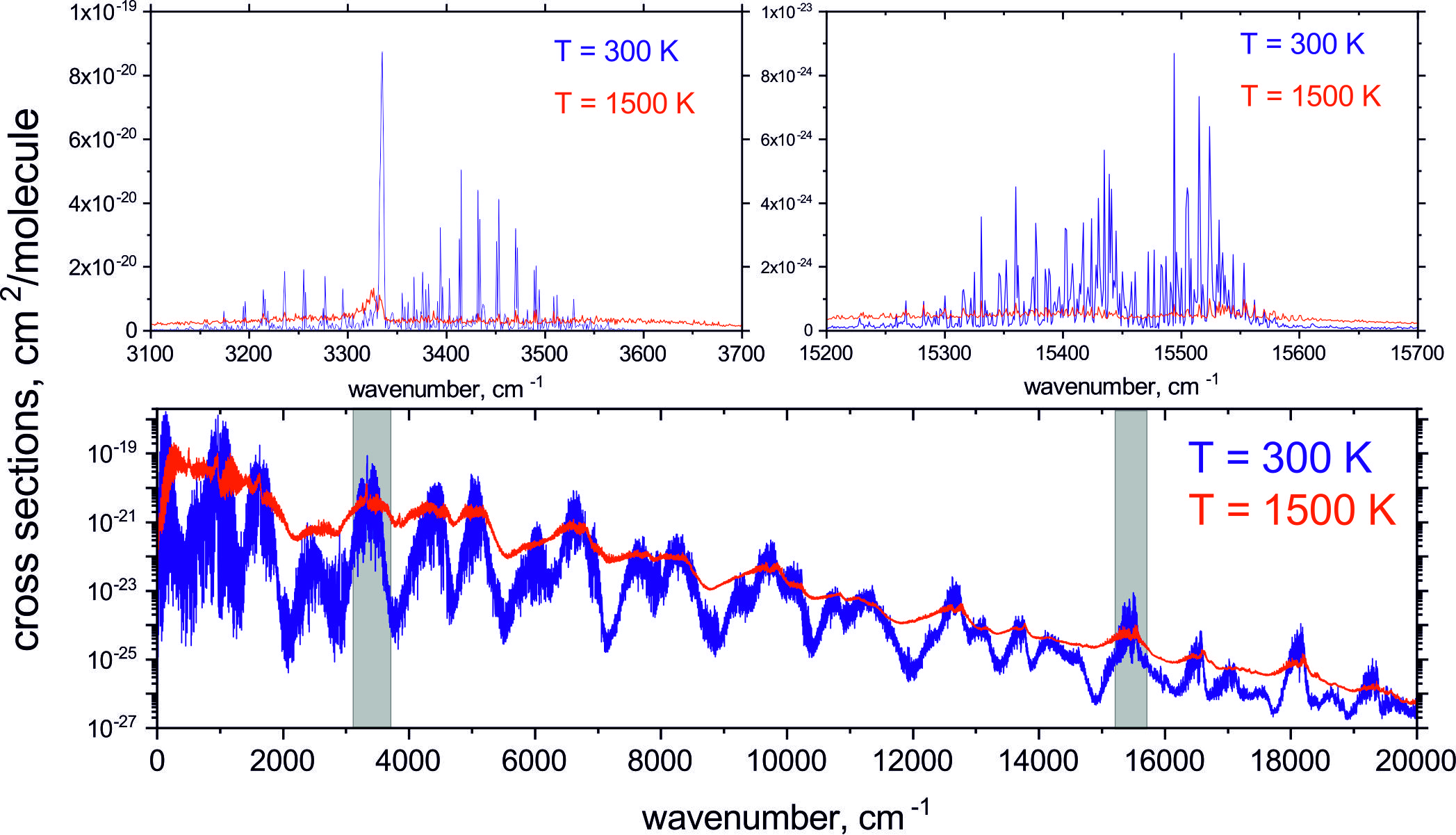}
\end{center}
\caption{Absorption cross sections generated using the ExoMol line list CoYuTe for  spectrum of NH$_3$ at $T=300$~K and 2000~K, with the Doppler profile on a wavenumber grid of 1~\cm. The cross sections were computed using the ExoCross program, which is developed to work with two file line list structure adopted by ExoMol.}
\label{fig:NH3}
\end{figure}

A new ExoMol line list for ammonia called CoYuTe \cite{jt771} has recently been
completed. CoYuTe, which also uses MARVEL energy levels \cite{jt608}, is both more
accurate and more extensive than the BYTe line list \cite{jt500} it replaces.
In particular, CoYuTe was found to provide a good model of the Jovian optical absorption
spectrum due to  ammonia \cite{jt745} although with a shift in positions of the main bands.
More laboratory work or analysis of existing laboratory work on the visible spectrum of ammonia
will be required to resolve this issue. An illustration of the ammonia absorption cross sections computed using CoYuTe is given in Fig.~\ref{fig:NH3}. 

\subsubsection{PH$_3$, Paper VII}

No change. \citet{20SoSaRa.PH3} use the SAlTY  PH$_3$ line list to assess the detectability of a possible PH$_3$  biosignature. A MARVEL project on the \ce{PH3} experimental data is under way, which will improve the quality of the line positions in SAlTY. 

\subsubsection{SO$_3$, Paper XVII}

There is a new UYT2 line list for SO$_3$ \cite{jt641}.

\subsection{Pentatomics}

\subsubsection{CH$_3$Cl, Paper XXIX}

There are new OYT line lists for methyl chloride \cite{jt733} covering both major
isotopes of chlorine.

\subsubsection{CH$_4$, Paper IV}

The ExoMol  YT10to10 line list \cite{jt564} has been extended to higher temperatures
\cite{jt698} to give the YT34to10 line list. However, at present the TheoReTS line list
of \citet{14ReNiTy.CH4} is the most accurate methane line list available and is
recommended for detailed studies. This line list forms the basis of the very recent
update to HITEMP \cite{20HaGoRe.CH4}.

The importance of methane and the huge number of lines required to accurately represent the opacity of hot 
methane \cite{jt572} has led to this
system being the primary focus of studies aimed at compacting these lines into a more
manageable form \cite{jt698,15HaBeBa.CH4,17ReNiTyi}. This issue is discussed in section~\ref{sect:data}.

\subsubsection{HNO$_3$, Paper XI}

No change. We note that, like H$_2$O$_2$, because of difficulties with assigning spectra even at room temperature,
HITRAN is very incomplete for 
nitric acid. The AlJS line list  gives  complete coverage at infrared wavelengths.

\subsubsection{SiH$_4$, Paper XXII}

There is a new OY2T line list for silane \cite{jt701}.

\subsection{Larger molecules}

\subsubsection{C$_2$H$_4$, Paper XXVII}

There is a new MaYTY line list for ethylene \cite{jt729}.

\subsection{MoLLIST species}

A number of diatomic species have line lists provided by the MoLLIST website \cite{MOLLIST}. These have been incorporated
into the ExoMol database, which now includes empirically-derived line lists for the following species \cite{jt790}:
CH \cite{14MaPlVa.CH};
NH \cite{14BrBeWe.NH,15BrBeWe.NH,18FeBeHo.NH};
OH \cite{16BrBeWe.OH}; 
AlCl \cite{18YoBexx.AlF};
AlF \cite{18YoBexx.AlF};
OH$^+$ \cite{17HoBexx.OH+,18HoBiBe.OH+};
CaF \cite{18HoBexx.CaF};
MgF \cite{17HoBexx.MgF};
KF and NaF \cite{16FrBeBr.NaF};
LiCl and LiF \cite{18BiBexx.LiF};
MgH \cite{13GhShBe.MgH};
TiH \cite{05BuDuBa.TiH};
CrH \cite{06ChMeRi.CrH};
FeH \cite{10WEReSe.FeH};
CP \cite{14RaBrWe.CP};
CN \cite{14BrRaWe.CN};
and CaH \cite{12LiHaRa.CaH,13ShRaBe.CaH}.
See Table \ref{tab:otherdata} for further details.

\subsection{HITRAN}

In general the line lists contained in the HITRAN database \citep{jt691} do not provide the temperature coverage needed 
for astrophysical applications.
However, for a few diatomic species very extensive line lists have been constructed. The line lists presented by
\citet{11LiGoBe.HCl} for HCl and \citet{15LiGoRo.CO} for CO are both valid over an extended temperature range and have been included in the ExoMol database.

The HF line list from \citet{13LiGoHa.HCl} and \citet{15CoHaxx.HF} is also adopted directly from HITRAN.

\subsection{N$_2$}
The empirical WCCRMT nitrogen molecule line list of \citet{18WeCaCr.N2} is somewhat unusual in that it only considers 
transitions between excited, triplet states of the molecule. However, N$_2$ is an important molecule with very limited spectral
signature at long wavelengths and WCCRMT may prove useful in hot or irradiated, usually non-LTE environments. For the partition function of \ce{N2}  the ground electronic state data from TIPS \citep{jt692s} were used.

\subsection{HD, HD$^+$, HeH$^+$}

One and two electron diatomics are important primordial species and may well play a role elsewhere. Recently \citet{19AmDiJo} created
new line lists using high level {\it ab initio} procedures. These line lists have been included in the database with the name 
ADSJAAM. In particular the ADSJAAM line lists for HeH$^+$ and HD supersedes those due to \citet{jt347} and \citet{jt506}, respectively.

\subsection{H$_2$}

A new external line list RACPPK by \citet{19RoAbCz.H2} for the ground electronic state of H$_2$ was converted to the ExoMol format. The line list contains a combination of electric quadrupole and magnetic dipole transitions computed from first principles.

\subsection{Other {\it ab initio} line lists}

A number of other {\it ab initio} line lists are also available. In general
{\it ab initio} line lists are only accurate for systems with very few electrons. The CLT line lists for LiH and LiH$^+$ due to
\citet{jt506} fall into the few-electron category and should be reliable. The other {\it ab initio} line lists provided, namely those for
ScH \cite{jt599},
CH$_3$F \cite{19OwYaKu.CH3F},
AsH$_3$ \cite{jt751},
P$_2$H$_2$ \cite{19OwYuxx.P2H2},
PF$_3$ \cite{jt752},
CH$_3$ \cite{19AdYaYu.CH3} and
YO \cite{jt774} 
must be regarded as intrinsically less accurate than other line lists provided by the database.
In the case of YO, work is in progress aimed at producing an empirical line list based on available experimental data. 
P$_2$H$_2$ occurs as two distinct isomers, the cis and trans forms. Separate line lists are provided for cis-P$_2$H$_2$ and trans-P$_2$H$_2$.

\subsection{SiH$_2$}

A new line list for SiH$_2$ \cite{jt779}  has been recently constructed in its ground electronic state using the standard ExoMol technology \cite{jt626}. The set of experimental data or their quality were, however, very limited which may affect the quality of the hot bands, which basically has to rely on the quality of the {\it ab initio} PES.

\subsection{NiH}

Empirical line lists for three isotopologues of NiH ($^{58}$NiH, $^{60}$NiH, $^{62}$NiH) known as HRV. These line lists
were constructed by \citet{09VaRiCr.NiH} and \citet{13HaRiTo.NiH}. However, at present these line lists are not available in ExoMol
format or accessible through the API.

\subsection{Partition functions}

The temperature dependent partition functions for most of the  ExoMol line lists are computed using the corresponding energy levels as collected in the States files as 
\begin{equation}
Q(T) = \sum_{i} g_{\rm ns}^{(i)} J_i(J_i + 1)  \exp\left(-\frac{c_2 \tilde{E}_i}{T}\right),
\end{equation}
where $g_{\rm ns}^{(i)}$ is the state dependent nuclear statistical weight, $J_i$ is the total angular momentum of the state $i$, $c_2$ is the second radiation constant (K\,cm) and $\tilde{E}_i$  is the corresponding energy term value (\cm). ExoMol uses the HITRAN convention \citep{06SiJaRo.method}, where the entire, integer $g_i = g_{\rm ns}^{(i)} J_i(J_i + 1)$ factors are explicitly included. 

Some of the line lists in the ExoMol database do not provide  sufficiently  complete sets  states for a proper evaluation of the molecular partition functions. For example, most of the MoLLIST line lists are constructed from or based on the  measured data only and therefore can be severely incomplete. In these cases the partition functions are either taken from external sources, such as the TIPS database \citep{jt692s}, generated using the empirical expansions, such as by \citet{81Irwin,84SaTaxx.partfunc,88Irwin,16BaCoxx.partfunc} or extrapolated using simplified models \citep{jt571}. The external partition functions often include the $g_{\rm ns}^{(i)}$ factors in the astrophysical convention (see, e.g. \citet{jt777}) and therefore have to be transformed to the HITRAN convention.

We strongly recommend that users use these partition functions rather than attempting to compute their own using
the ExoMol States files. These States files are not constructed with a view to delivering reliable partition functions
and in a number of cases use of them has been found to lead to problems.

\subsection{VUV sections}

A section is provided based on measured ultraviolet cross sections for key species. For many molecules their ultraviolet spectrum is a mixture of line and quasi-continuum absorption which cannot be represented as a line list. The new VUV section will provide 
temperature-dependent absorption cross sections. Recent results suggest the VUV absorption by H$_{2}$O enhances the production of OH, which plays an important molecule in atmospheric chemistry of exoplanetary atmospheres \cite{20RaScHa.H2O}.  Currently, the VUV cross sections are provided for H$_2$O, H$_2$, CO$_2$, SO$_2$, NH$_3$, H$_2$CO and C$_2$H$_4$ measured by Fateev's lab at the Danish Technical University (DTU) \cite{20FaClYi} and for \ce{CO2} by \citet{18VebeFa.CO2}. The temperature and wavelength coverage is illustrated in Table~\ref{t:VUV}.

The cross sections are given in the common two-column ASCII format separated by spaces, where the first column contains the wavelength in nm and the second column contains the absorption cross sections in cm$^{2}/$molecule using Fortran format: \texttt{(F10.3,1x,E13.6)}. The VUV cross section file names have the following structure \\
\texttt{\textquotesingle{}<ISOTOPOLOGUE>\_\_<DATASET>\_\_<RANGE>\_\_T<TEMP>K\_\_P<PRESSURE>bar\_\_<STEP>.nm\textquotesingle{}}, where \texttt{ISOTOPOLOGUE} is the isotopologue name, \texttt{DATASET} is the name of the line list, \texttt{RANGE} is the wavelength range in nm, \texttt{TEMP} is the temperature in K, \texttt{PRESSURE} is the pressure in bar, \texttt{STEP} is the wavelength step in nm. For example, the States file of the VUV line list for \ce{CO2} the filename:\\  \url{12C-12O2__Venot-2018__116.90-230.00__T0800K__P0bar__0.03.nm}.


\begin{table}[H]
    \caption{VUV absorption cross sections on ExoMol, grid spacing 0.01--0.015~nm, presented in natural abundance.}
    \label{t:VUV}
    \centering
    \begin{tabular}{lrr}
    \hline
    \hline
    Molecule & Temperature (K) & Range (nm) \\
    \hline
    \multicolumn{3}{c}{DTU data}\\
    \hline
         \ce{H2O} & 423 &  110--230  \\
         \ce{H2O} & 573 &  110--230  \\
         \ce{H2O} & 1630 &  182--237  \\
         \ce{H2O} & 1773 &  182--237  \\
         \ce{CO2} & 1160 & 109--324  \\
         \ce{SO2} & 423 & 110--230  \\
         \ce{NH3} & 289 & 113--201  \\
         \ce{NH3} & 296 & 113--201  \\
         \ce{H2CO} & 303 & 110--230  \\
         \ce{H2CO} & 353 & 110--230  \\
         \ce{H2CO} & 423 & 110--230  \\
         \ce{H2CO} & 573 & 110--230  \\
         \ce{C2H4} & 562 & 113--201  \\
\hline 
    \multicolumn{3}{c}{\protect\citet{18VebeFa.CO2}}\\
\hline
         \ce{CO2} & 170--800~K & 115-220 \\
    \hline
    \hline
    \end{tabular}
\end{table}


\section{Data Provided}

Table~\ref{tab:datsum} provides a summary of different types of data provided. The website
provides two routes to accessing these data. Users can search by molecule which will show all 
the types of data available for each isotopologues. Alternatively it is possible to search by data
type in which case a list of molecules for which data of the specified type is available.
The following section lists the data types and their file name extensions for the various data and metadata provided by ExoMol. A more technical specification of the data structures and how to access data using the application program interface (API) is provided in Section~\ref{sect:data-formats-and-api}.

\begin{table}[H]
    \centering
    \caption{Summary of data provided by the ExoMol Database}
    \label{tab:datsum}
    \begin{tabular}{ll}
\hline\hline
Data type 
\\
\hline
Line lists \\
Absorption cross sections \\
VUV absorption cross sections \\
Pressure broadening coefficients \\
Temperature dependent super-lines (histograms) \\
Partition functions  \\
Cooling functions \\
Specific heat - heat capacity \\
Examples of ExoCross input files \\
Temperature and pressure dependent opacities \\
Spectroscopic Models    \\
\hline\hline
\end{tabular}
\end{table}

\subsection{Data Structure}

\begin{table}[H]
\caption{Specification of the ExoMol file types. (Contents in brackets are
optional.)}
\label{tab:files}
\tiny
\begin{tabular}{lcll}
\hline\hline
File extension & $N_{\rm files}$&File DSname &  Contents\\
\hline
\texttt{.all} &1& Master& Single file defining contents of the ExoMol
database..\\
\texttt{.def} &$N_{\rm tot}$& Definition& Defines contents of other files for
each isotopologue.\\
\texttt{.states} &$N_{\rm tot}$& States & Energy levels, quantum numbers,
lifetimes, (Land\'e $g$-factors, Uncertainties).\\
\texttt{.trans} &$^a$& Transitions & Einstein $A$ coefficients, (wavenumber).\\
\texttt{.broad} &$N_{\rm mol}$& Broadening & Parameters for pressure-dependent
line profiles.\\
\texttt{.cross}&$^b$& Cross sections& Temperature or temperature and pressure-dependent
cross sections.\\
\texttt{.kcoef}&$^c$& $k$-coefficients& Temperature and pressure-dependent
$k$-coefficients.\\
\texttt{.pf}& $N_{\rm tot}$&Partition function&   Temperature-dependent
partition function,
(cooling function).\\
\texttt{.dipoles}&$N_{\rm tot}$& Dipoles & Transition dipoles including
phases.\\
\texttt{.super} &$^d$& Super-lines & Temperature dependent super-lines (hisograms) on a wavenumber grid.\\
\texttt{.nm} &$^e$ & VUV cross sections& Temperature and pressure dependent VUV cross-sections (wavelength, nm).\\
\texttt{.fits}, \texttt{.h5}, \texttt{.kta}  &$^f$ & Opacities & Temperature and pressure dependent  opacitities for radiative-transfer applications.\\
\hline
\texttt{.overview}&$N_{\rm mol}$& Overview & Overview of datasets available.\\
\texttt{.readme}&$N_{\rm iso}$& Readme & Specifies data formats.\\
\texttt{.model}&$N_{\rm iso}$& Model & Model specification.\\
\hline\hline
\end{tabular}

\noindent
$N_{\rm files}$ total number of possible files;\\
$N_{\rm mol}$ Number of molecules in the database;\\
$N_{\rm tot}$ is the sum of $N_{\rm iso}$ for the $N_{\rm mol}$ molecules in the
database;\\
$N_{\rm iso}$ Number of isotopologues considered for the given molecule.\\
$^a$ There are $N_{\rm tot}$ sets of \texttt{.trans} files but for molecules
with large
numbers of transitions the \texttt{.trans} files are subdivided into wavenumber
regions.\\
$^b$ There are $N_{\rm cross}$ sets of \texttt{.cross} files for
isotopologue.\\
$^c$ There are $N_{\rm kcoef}$ sets of \texttt{.kcoef} files for  each
isotopoplogue.\\
$^d$ There are $N_T$ sets of $T$-dependent super-lines. 
$^e$ There are $N_{VUV}$ sets of VUV cross sections.
$^f$ Set of opacity files in in the format  native  to specific radiative-transfer programs.

\end{table}

The general ExoMol approach is molecule-by-molecule: 
 a comprehensive line list is created for a particular molecule
 and made available in the database. The line list for each 
isotopologue is stored as a separate data structure which can be accessed directly or via the 
application program interface (API) described in the following sections. 
Table~\ref{tab:files} specifies the file types that can be available for each isotopologue. The \texttt{.states} and \texttt{.trans}
files are the heart of the ExoMol data structure \cite{jt548} and define what 
has become known as the \lq\lq ExoMol format\rq\rq. These files are available for all isotopologues; other files may not be.
ExoMol format files can be used and provided by  the effective Hamiltonian code PGOPHER
\cite{PGOPHER}.

A manual providing technical specifications of the 
database is included as supplementary data to this article and is included on the website where it will be updated as the 
data base evolves.
Only the relatively few changes that have been implemented since the 2016 release are discussed below.

\subsection{Super-lines}

Super-lines \citep{TheoReTS,jt698} represent a novel, compact way of storing the opacity data. Super-lines are constructed as temperature-dependent intensity histograms by summing all absorption coefficients within a wavenumber bin   centred around a grid point $\tilde\nu_k$.  For each $\tilde\nu_k$ the total absorption intensity ${I}_k(T)$ is computed as a sum of absorption line intensities $I_{if}$ from all $i\to f$ transitions   falling into the wavenumber bin $[\tilde\nu_k-\Delta \tilde{\nu}_k/2 \ldots \tilde\nu_k+\Delta \tilde{\nu}_k/2 ] $ at the given temperature $T$. Each grid point $\tilde\nu_k$ is then treated as a line position of an artificial transition   (super-line) with an effective absorption intensity ${I}_k(T)$. 
The number of data points in the super-lines can be drastically reduced without  significant loss of accuracy when computing pressure-dependent cross sections and can be combined with standard, pressure, temperature and frequency dependent line profiles. The super-lines cannot be associated with any specific upper/lower states and therefore the line broadening parameters  used cannot  depend on quantum numbers. 

The super-line list files have the format of cross sections represented by two columns, wavenumbers (\cm) and super-line absorption intensity (cm$/$molecule). The Fortran format is \texttt{(F12.6,1x,ES14.8)}. The super-lines are computed on a grid of temperatures from 100~K to $T_{\rm max}$. For the resolution an adaptive grid of $R= 1\,000\,000$ is used.  For the technical details of super-line construction and numerical tests see \citet{jt698}. 

Super-lines are currently available for \ce{H2O}, \ce{NH3}, \ce{CH3Cl}, \ce{C2H2} and \ce{C2H4}. 


\subsection{Specific heat}

\begin{figure}[t]
\begin{center}\includegraphics[width=0.75\textwidth]{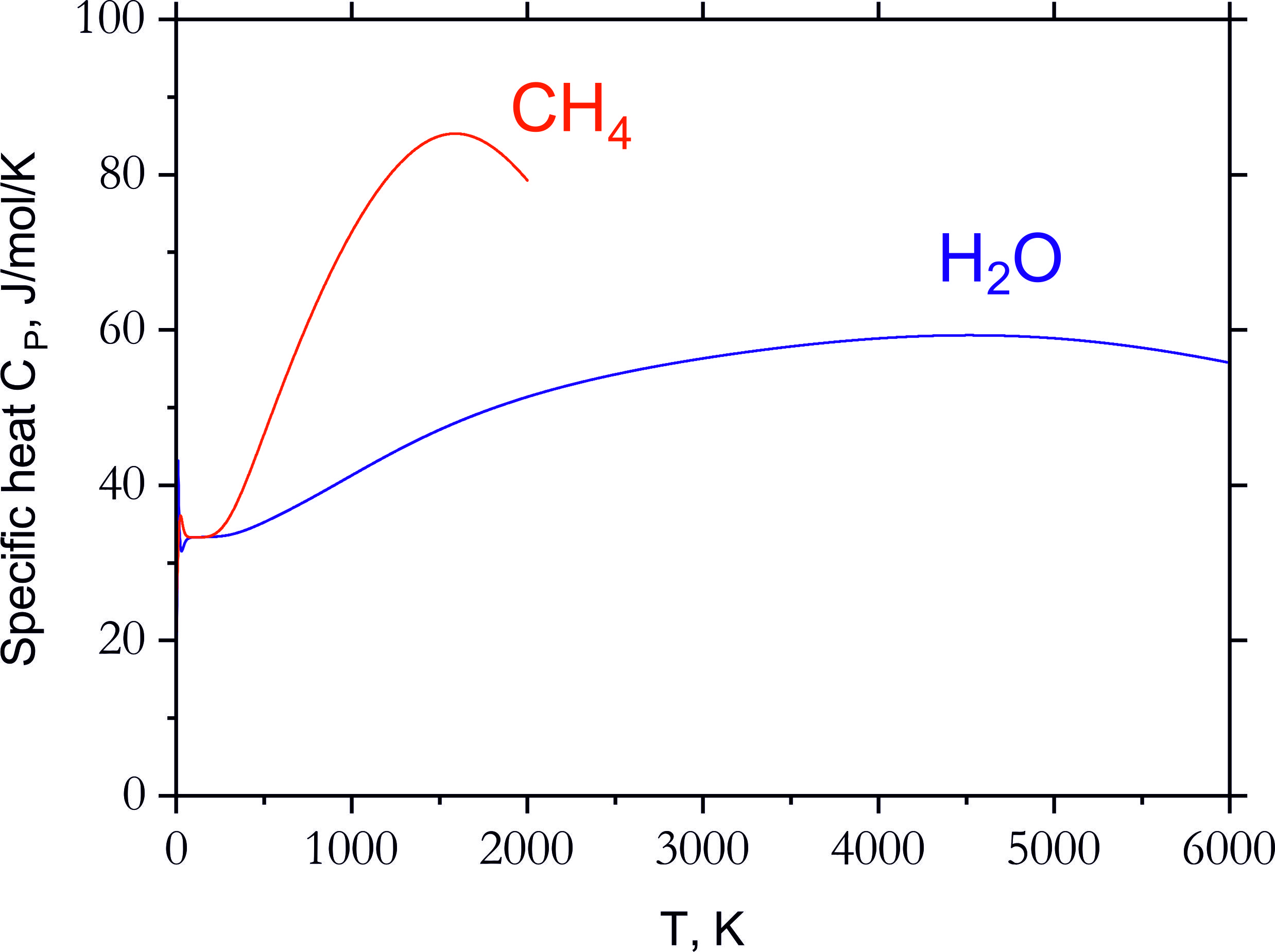}
\end{center}
\caption{Specific heat at constant pressure of H$_2$O as computed by \citet{jt661} and CH$_4$  generated   the YT10to10 ro-vibrational  energies \cite{jt564}.}
\label{fig:Cp}
\end{figure}

The specific heat at constant pressure $C_p$ is computed on a grid of temperatures of 1~K from the molecule partition function as given by \citep{jt661}
\begin{equation}
C_p(T) = R \left[\frac{Q''}{Q}-\left(\frac{Q'}{Q}\right)^2 \right] + \frac{5R}{2},
\end{equation}
where the second term is the translational contribution, 
$$
Q'(T) = T\frac{d Q}{d T},
$$
\begin{equation}
Q''(T) = T^2\frac{d^2 Q}{d T^2} + 2Q'
\end{equation}
and $R$ is the gas constant.  The specific heat values are given in units of J mol$^{-1}$ K$^{-1}$ and are currently provided for \ce{H2O} (taken from \citet{jt661})   and \ce{CH4} (computed using the YT10to10 ro-vibrational energies \cite{jt564}) only (see Fig.~\ref{fig:Cp}). 

\subsection{Line shifts using the diet format}

For high resolution applications it will be important to take into account the pressure dependent line shifts. To this end a line shift diet has been introduced using the same data structure used for pressure dependent line broadening diet \citep{jt684}. An example for \ce{H2O} is given in Table~\ref{t:shift}.

\SaveVerb{term}|1H-16O2__a0.shift|
\begin{table}[H]
\caption{File \protect\UseVerb{term} A pressure line  \texttt{.shift} file for
H$_2$O: portion of the file (upper part); field specification (lower part).}
\label{t:shift} \footnotesize
\begin{center}
\begin{tabular}{lll}
\hline\hline
Code & Shift & J \\
\hline
a0 & 0.0001 &  0 \\
a0 & 0.0001 &  1 \\
a0 & 0.0001 &  2 \\
a0 & 0.0001 &  3 \\
a0 & 0.0001 &  4 \\
\ldots & \ldots & \ldots \\
\end{tabular}
\begin{tabular}{llll}
\hline
Field & Fortran Format & C Format & Description \\
\hline
code & A2 & $\%$2s & Code identifying quantum number set following $J\pp$* \\
$\gamma_{\textrm{ref}}$ & F15.6 & \%15.6f & Line shift at reference
temperature and pressure in \cm\ \\
$J\pp$ & I7/F7.1 & $\%$7d & Lower $J$-quantum number \\
\hline
\end{tabular}
\end{center}
\noindent
*Code definition:
a0 = none
\end{table}

\subsection{New broadening parameters}

The pressure broadening parameters are provided for 15 molecules  and are listed in Table~\ref{t:pressure:diet} using the pressure-broadening diet \citep{jt684}.

The \texttt{.broad} file has a hierarchical structure; each record starts with four compulsory columns: a label defining the broadening scheme (`a0', `a1', \ldots), values of 
$\gamma_{\textrm{ref}}$, $n$ and $J''$. The compulsory fields are followed by additional quantum numbers when a more detailed specification of the quantum assignments is available. The most basic  scheme `a0' represents broadening with $J\pp$ dependence only. The  additional basic scheme `a1' is used for the case with the $(J',J'')$  dependence. Any other schemes (e.g. `a2', `a3', `a4', `b1', `b2', `c1', `c2') are the molecule specific and should be described as part of the line list specifications. The broadening data format is illustrated in Table~\ref{tab:broad:a0:a1:a2}. 

The requirement for enhanced line broadening parameters was identified by \citet{jt773} is their study on the need  for laboratory data requirements for studies  of exoplanetary atmospheres. Recognising this, our plan is significantly enhance the treatment
pressure effects in future releases of the ExoMol database.

\begin{table}[H]
\centering
    \caption{Pressure broadening parameters in ExoMol}
    \label{t:pressure:diet}
\begin{tabular}{lll}
\hline\hline
Molecule & Broadener & Scheme \\
\hline
CS & air, self & a0  \\
HCl & \ce{H2}, He, self, air, \ce{CO2} & a0 \\
HF & \ce{H2}, He & a0 \\
CO & \ce{H2}, He & a0 \\ 
NO & air & a0 \\
\ce{H2O} & \ce{H2}, He, self, air & a0,a1 \\
\ce{CO2} & air, self & a0  \\
\ce{SO2} & \ce{H2}, He & a0, a5 \\ 
HCN & \ce{He}, He & a0 \\ 
OCS & \ce{He}, He & a0 \\ 
\ce{PH3} & \ce{He}, He & a0, c1 \\
\ce{NH3}  & \ce{He}, He & a0 \\ 
\ce{H2CO} & \ce{He}, He & a0 \\ 
\ce{C2H2} & \ce{He}, He & a0 \\ 
\ce{CH4}& \ce{He}, He, \ce{CO2} & a0, a1 \\  
\hline
\hline
\end{tabular}
\end{table}

\begin{table}[H]
\caption{Example of the three basic  broadening schemes in the ExoMol diet, `a0', `a1', `a2'.}
\label{tab:broad:a0:a1:a2} \footnotesize
\begin{center}
\begin{tabular}{crrrrr}
\hline
\hline
Label & $\gamma$ & $n$ & $J''$ & $J'$ & $K$ \\
\hline
a0 & 0.0860 & 0.096 & 0 \\
a0 & 0.0850 & 0.093 & 1 \\
... & & & \\
a1 & 0.0860 & 0.096 & 0 & 1  \\
a1 & 0.0850 & 0.093 & 1 & 2  \\
... & & & \\
\end{tabular}
\end{center}
\end{table}

\subsection{ExoMolOP: Opacities}

Recently \citet{jt801} computed
opacity cross sections and $k$-tables for all molecules available from the ExoMol database plus  some atomic data from NIST. 
These data are formatted for use in various retrieval codes including Tau-REx \cite{jt593,jt611,TauRexIII}, ARCiS \cite{19MiOrCh.arcis},
petitRADTRANS \cite{19MoWaBo.petitRADTRANS}, and NEMESIS \cite{NEMESIS}.
Data were calculated on temperature-pressure grids suitable for characterising a variety of exoplanet and stellar
atmospheres. Broadening parameters were taken from the literature where available, with broadening parameters used for a
known molecule with a similar dipole moment, where they are not available. Tables of cross sections and $k$-coefficients are provided
on ExoMol  as part of the line list webpage. The format of the data and the resolution (grid spacing) is application dependent, for example the opacities for  ARCiS, petitRADTRANS, and NEMESIS are given as $k$-tables at the resolving power of $R$= 1000 while  Tau-REx works with cross-sections with $R = 15\,000$. 
 
The opacity cross sections and $k$-tables will also be made available via the virtual atomic and molecular data centre (VAMDC) portal
\cite{jt481,jt630}. We note that ExoMol data has also been extensively used to construct the EXOPLINES molecular absorption cross-Section database for brown dwarf and giant exoplanet atmospheres \cite{exoplines}.

\subsection{Post-processing}\label{s:exocross}

ExoMol provides post-processing capabilities through the program ExoCross \cite{jt708}.
ExoCross has many functions such as generating pressure and temperature dependent cross sections, partition functions, specific heat,  state-resolved radiative lifetimes, non-LTE spectra, electric dipole, electric quadrupole and magnetic dipole spectra. ExoCross can read data in both ExoMol and HITRAN \cite{jt557} formats and output it in these formats as well as SPECTRA (\url{http://spectra.iao.ru/}) and Phoenix formats \cite{PHOENIX}. Should data be needed in say HITRAN format, it is strongly recommended that the data is downloaded to a local computer in the much more compact ExoMol format and then processed using ExoCross.
Examples of ExoCross input files are provided on the website.

There are also a number of Python utilities available on the ExoMol website.
These are now largely redundant as ExoCross provides all functionality required to work with ExoMol line lists.
However, we mention them for completeness:
Utility \texttt{extract\_\_trans.py}
reads the \texttt{.trans} in \texttt{.bz2} format
without requiring it to be uncompressed. The \texttt{ExoMol\_to\_HITRAN.py} script converts ExoMol format
to HITRAN format; this should be used with caution as ExoMol format is significantly more compact
than HITRAN format and the line list files are large. The program \texttt{exomol2gf.py}
can be use to generate oscillator strengths.
 
\subsection{New web services}

New web services include:
\begin{itemize}
    \item A molecule and line list search. 
    \item A graphical illustration of a line list represented by absorption spectra at two  or three temperatures (e.g. 300~K, 2000~K and 5000~K) computed using a Doppler line profile on a grid of 1~\cm.
    \item The ExoMol bibliography in the BibTeX format is stored at and version controlled by GitHub at \url{https://github.com/ExoMol/bib}.
\end{itemize}

\section{ExoMol data formats and API}\label{sect:data-formats-and-api}

\subsection{Format of the \texttt{states} and \texttt{trans} files}

 The formats of the States and Transitions files are  specified in Table~\ref{t:states:format} and \ref{t:trans:format}.  Tables~\ref{t:states} and \ref{t:trans} show typical examples of the States (.states) and Transitions (.trans)  files.
A significant new feature of the 2020 update is the uncertainty field (\cm) in the States file appearing as column 5 after the $J$ values; this currently an optional feature but will become a compulsory column  for the ExoMol States file. The uncertainty values typically come from three different sources: (i) the uncertainties of the MARVEL energies ($\sim$0.000001--0.1~\cm),  (ii) uncertainties obtained as the fitting observed -- calculated error of energies as part of the refinement to experimental data (energies or line positions)  participated in the refinement ($\sim$ 0.001 -- 1~\cm) and (iii) roughly estimated uncertainties for all other states that cannot be verified against existing experiment ($\sim$ 0.1 -- 10 \cm).  
The uncertainty in the States file are given using the same format as the energy term values, i.e. with six  decimal places after the decimal point, see Table~\ref{t:states:format}. 

To date the new standard of the States file has being applied to the only a number of key/ recent line lists  namely \ce{H2O} (POKAZATEL) \cite{jt734}, AlH (AlHambra) \cite{jt732}, C$_2$ {8states} \cite{jt736,jt809}, HCCH (aCeTY) \citep{jt780}, \ce{CO2} (UCL-4000) \citep{jt804}, \ce{H3O+} (eXeL) \citep{jt805} and TiO (Toto) \cite{jt760}. A rolling programme is in place for updating the 
other files so the uncertainty is uniformly available for all sources.



\begin{table}[H]
\caption{Updated specification of the ExoMol States file.}
\label{t:states:format}
\begin{tabular}{llll} \hline
Field & Fortran Format & C Format & Description\\ \hline
$i$ & \texttt{I12} & \texttt{\%12d} & State ID\\
$E$ & \texttt{F12.6} & \texttt{\%12.6f} & State energy in $\mathrm{cm^{-1}}$\\
$g_\mathrm{tot}$ & \texttt{I6} & \texttt{\%6d} & State degeneracy\\
$J$ & \texttt{I7/F7.1} & \texttt{\%7d/\%7.1f} & $J$-quantum number
(integer/half-integer)\\
($\Delta E$) & \texttt{F12.6} & \texttt{\%12.6f} & Energy uncertainty in $\mathrm{cm^{-1}}$ (currently optional)\\
($\tau$) & \texttt{ES12.4} & \texttt{\%12.4e} & Lifetime in s (optional)\\
($g$) & \texttt{F10.6} & \texttt{\%10.6f} & Land\'e $g$-factor  (optional)\\
(Extra) & - & -& Extra quantum numbers, any format (optional)\\
\hline
\end{tabular}
\end{table}

\begin{table}[H]
\caption{Specification of the Transitions file.}
\label{t:trans:format}
\begin{tabular}{llll} \hline
Field & Fortran Format & C Format & Description\\ \hline
$i$ & \texttt{I12} & \texttt{\%12d} & Upper state ID\\
$f$ & \texttt{I12} & \texttt{\%12d} & Lower state ID\\
$A$ & \texttt{ES10.4} & \texttt{\%10.4e} & Einstein $A$ coefficient in
$\mathrm{s^{-1}}$ \\
$\tilde{\nu}_{fi}$&\texttt{E15.6} & \texttt{\%15.6e} & Transition wavenumber in
cm$^{-1}$ (optional). \\
 \hline
\end{tabular}
\end{table}

\begin{table}[H]
{\footnotesize
\caption{An example of an extract from the final States file for UCL-4000 of \ce{CO2} \citep{jt804}. }
\label{t:states}
\tabcolsep=5pt
\begin{tabular}{rrrrrccrrrr}
\hline\hline
$i$ & \multicolumn{1}{c}{$\tilde{E}$} & $g_{\rm tot}$  & $J$ & \multicolumn{1}{c}{unc.} & $\Gamma$ &$e/f$ &  $n_1$ &$n_2^{\rm lin}$ & $l_2$ &$n_3$\\
\hline
           1&     0.000000&      1&       0&       0.000500& A1&  e &   0 &  0 &  0 &  0\\
           2&  1285.408200&      1&       0&       0.000500& A1&  e &   0 &  2 &  0 &  0\\
           3&  1388.184200&      1&       0&       0.005000& A1&  e &   1 &  0 &  0 &  0\\
           4&  2548.366700&      1&       0&       0.000500& A1&  e &   1 &  2 &  0 &  0\\
           5&  2671.142957&      1&       0&       0.005000& A1&  e &   2 &  0 &  0 &  0\\
           6&  2797.136000&      1&       0&       0.005000& A1&  e &   1 &  2 &  0 &  0\\
           7&  3792.681898&      1&       0&       0.005000& A1&  e &   1 &  4 &  0 &  0\\
           8&  3942.541358&      1&       0&       0.005000& A1&  e &   3 &  0 &  0 &  0\\
           9&  4064.274256&      1&       0&       0.005000& A1&  e &   3 &  0 &  0 &  0\\
          10&  4225.096148&      1&       0&       0.005000& A1&  e &   1 &  4 &  0 &  0\\
          11&  4673.325200&      1&       0&       0.000500& A1&  e &   0 &  0 &  0 &  2\\
          12&  5022.349428&      1&       0&       0.005000& A1&  e &   1 &  6 &  0 &  0\\
          13&  5197.252900&      1&       0&       0.005000& A1&  e &   3 &  2 &  0 &  0\\
          14&  5329.645446&      1&       0&       0.005000& A1&  e &   4 &  0 &  0 &  0\\
          15&  5475.553054&      1&       0&       0.000500& A1&  e &   3 &  2 &  0 &  0\\
          16&  5667.644584&      1&       0&       0.005000& A1&  e &   2 &  4 &  0 &  0\\
          17&  5915.212302&      1&       0&       0.000500& A1&  e &   0 &  2 &  0 &  2\\

\hline\hline
\end{tabular}

\mbox{}\\

{\flushleft
$i$:   State counting number.     \\
$\tilde{E}$: State energy in \cm. \\
$g_{\rm tot}$: Total state degeneracy.\\
$J$: Total angular momentum.            \\
unc.: Uncertainty in \cm.}     \\
$\Gamma$:   Total symmetry index in \Cv{2}(M).\\
$e/f$: Kronig rotationless parity. \\
$n_1$: Normal mode stretching symmetry ($A_1$) quantum number. \\
$n_2^{\rm lin}$ : Normal mode linear molecule bending ($A_1$) quantum number. \\
$l_2$: Normal mode vibrational angular momentum quantum number. \\
$n_3$: Normal mode stretching asymmetric ($B_1$) quantum number. }
\end{table}

\begin{table}
\caption{  Extract from the Transitions file for CaO.}
\label{t:trans}
\begin{center}
\begin{tabular}{rrcc}
\hline\hline
$f$     &  $i$          &               $A_{fi}$ &  $\tilde{\nu}_{fi}$\\
\hline
       10571    &        10884   &    9.5518E-06       &        120.241863      \\
       21053    &        21375   &    1.9515E-05       &        120.242886      \\
        8726    &         9672   &    1.8658E-04       &        120.243522      \\
       11655    &        11950   &    5.0065E-06       &        120.243733      \\
       93209    &        93967   &    5.7055E-03       &        120.244192      \\
        2228    &         3175   &    7.3226E-07       &        120.244564      \\
       46727    &        46432   &    1.0599E-04       &        120.244658      \\
       44436    &        44774   &    1.4626E-04       &        120.245583      \\
       29037    &        28723   &    1.8052E-04       &        120.245669      \\
        4458    &         4805   &    1.0431E-08       &        120.246396      \\
       69313    &        68434   &    5.0531E-06       &        120.248178      \\
       22640    &        22985   &    1.1281E-07       &        120.248891      \\
       57027    &        56721   &    7.1064E-06       &        120.250180      \\
   \hline
\end{tabular}

\noindent
 $f$: Upper  state counting number;
$i$:  Lower  state counting number; $A_{fi}$:  Einstein-A
coefficient in s$^{-1}$; $\tilde{\nu}_{fi}$: transition wavenumber in \cm.

\end{center}
\end{table}

\subsection{Formats for other data types}

Table~\ref{tab:broad_format} shows the format of the \texttt{.broad} files containing the line broadening parameters. Table~\ref{tab:cross} shows the format of the cross section \texttt{.cross} and super-lines \texttt{.super} files. Tables~\ref{tab:pf} and \ref{tab:cp}  show the format of the partition function  \texttt{.pf} and specific heat \texttt{.cp} files.

\begin{table}[H]
\caption{Specification of the mandatory part of the pressure broadening
parameters file.}
\label{tab:broad_format} \footnotesize
\begin{center}
\begin{tabular}{llll}
\hline
Field & Fortran Format & C Format & Description \\
\hline
code & A2 & \%2s & Code identifying quantum number set following $J\pp$ \\
$\gamma_{\textrm{ref}}$ & F6.4 & \%6.4f & Lorentzian half-width at reference 
temperature and pressure in \cm/bar \\
$n$ & F6.3 & \%6.3f & Temperature exponent \\
$J\pp$ & I7/F7.1 & \%7d/\%7.1f & Lower $J$-quantum number integer/half-integer
\\
\hline
\end{tabular}
\end{center}

\noindent
Fortran format, $J$ integer: \texttt{(A2,1x,F6.4,1x,F6.3,1x,I7)}\\
or $J$ half-integer: \texttt{(A2,1x,F6.4,1x,F6.3,1x,F7.1)}\\
\end{table}

\begin{table}[H]
\caption{Specification of the \texttt{.cross} cross section and \texttt{.stick}  file
format}\label{tab:cross}
\begin{tabular}{llll}
\hline
Field & Fortran Format & C Format & Description\\
\hline
$\tilde{\nu}_i$ & \texttt{F12.6} & \texttt{\%12.6f} & Central bin wavenumber,
$\mathrm{cm^{-1}}$\\
$\sigma_i$ & \texttt{ES14.8} & \texttt{\%14.8e} & Absorption cross section,
$\mathrm{cm^2\,molec^{-1}}$\\
$\alpha_i$ & \texttt{ES14.8} & \texttt{\%14.8e} & Absorption coefficient,
$\mathrm{cm\,molec^{-1}}$\\
\hline
\end{tabular}

\noindent
Fortran format: \texttt{(F12.6,1x,ES14.8)}\\
\end{table}

\begin{table}[H]
\caption{Specification of the \texttt{.pf}  partition function file.}
\label{tab:pf}
\begin{tabular}{llll}
\hline
Field & Fortran Format & C Format & Description\\
\hline
$T$ & \texttt{F8.1} & \texttt{\%8.1d} & Temperature in K\\
$Q(T)$ & \texttt{F15.4} & \texttt{\%15.4d} & Partition function
(dimensionless).\\
$W(T)$ & \texttt{ES12.4} & \texttt{\%12.4e} & Cooling function in ergs s$^{-1}$
molecule$^{-1}$ (if available).\\
\hline
\end{tabular}

\noindent
Fortran format: \texttt{(F8.1,1x,F15.4,1x,ES12.4)} or \texttt{(F8.1,1x,F15.4)}
\\
\end{table}

\begin{table}[H]
\caption{Specification of the \texttt{.cp}  specific heat file.}
\label{tab:cp}
\begin{tabular}{llll}
\hline
Field & Fortran Format & C Format & Description\\
\hline
$T$ & \texttt{F8.1} & \texttt{\%8.1d} & Temperature in K\\
$C_p(T)$ & \texttt{F15.4} & \texttt{\%15.4d} & Specific heat function
(dimensionless).\\
\hline
\end{tabular}

\noindent
Fortran format:  \texttt{(F8.1,1x,F15.4)}
\\
\end{table}

\subsection{API}\label{sect:api}

\subsubsection{Searching for data through the API}\label{sect:api:1}

The molecules and isotopologues available in ExoMol are listed in a master file located at: \url{www.exomol.com/exomol.all}. Given a molecule or list of isotopologues, ExoMol can be searched for recommended datasets using the API which can be queried using the \texttt{HTTP GET} request method described below. The structure of a JSON file is illustrated in Fig.~\ref{fig:API:search}.

\begin{figure}[H]
    \centering
{
    \renewcommand{\baselinestretch}{0.9}
    \color{blue}
    \setlength{\parindent}{1em}
    \setlength{\parskip}{0em}
\begin{verbatim}
    {<ISOTOPOLOGUE>:
        {'<DATATYPE>':
            {'<DATASET>':
                'description': <DATASET-DESCRIPTION>,
                'files': [{'description': <FILE-DESCRIPTION>,
                           'url': <URL>,
                           'size': <SIZE>
                          },
                          ...
                         ]
            },
            {'<DATASET>':
                ...
            },
            ...
        },
        {'<DATATYPE>':
            ...
        },
        ...
    },
    {<ISOTOPOLOGUE>:
        {
        ...
        }
    }
\end{verbatim}
}
\caption{A JSON API structure used for search queries using the \texttt{HTTP GET} request in ExoMol.}
    \label{fig:API:search}
\end{figure}

To search for all data sets (and their files) related to a single molecule, use the query:
{
    \renewcommand{\baselinestretch}{0.9}
    \color{blue}
    \setlength{\parindent}{1em}
    \setlength{\parskip}{0em}
\begin{verbatim}
http://exomol.com/api?molecule=<MOLECULE>
\end{verbatim}
}

For example, the URL \url{http://exomol.com/api?molecule=NH3} returns a JSON data structure describing all files related to the ammonia molecule. Within this, the list of files belonging to the BYTe-15 line list for $\mathrm{^{15}N^{1}H_3}$ is accessible by traversing the JSON object with:
{
    \renewcommand{\baselinestretch}{0.9}
    \color{blue}
    \setlength{\parindent}{1em}
    \setlength{\parskip}{0em}
\begin{verbatim}
json_result['(15N)(1H)3']['linelist']['BYTe-15']['files']
\end{verbatim}
}
\noindent
and the URL for the States file belonging to this line list is at:
{
    \renewcommand{\baselinestretch}{0.5}
    \color{blue}
    \setlength{\parindent}{1em}
    \setlength{\parskip}{0em}
\begin{verbatim}jsonresult['(15N)(1H)3']['linelist']['BYTe-15']['files'][1]['url'].
\end{verbatim}
}
Note that molecules are identified by simple text strings (no subscript or superscript symbols) but that special characters must be URL-encoded: for example, the $\mathrm{HeH^+}$ cation is identified by \texttt{molecule=HeH\%2B}. Most software will provide this ``percent-encoding'' automatically.

Searches can be further refined by isotopologue by setting the query keyword \texttt{isotopologues} to a comma-delimited sequence; for example:
{
    \renewcommand{\baselinestretch}{0.9}
    \color{blue}
    \setlength{\parindent}{1em}
    \setlength{\parskip}{0em}
\begin{verbatim}
http://exomol.com/api/?isotopologues=(12C)(32S),(12C)(34S)
\end{verbatim}
}
\noindent
Restricting the search by datatype is also supported:
{
    \renewcommand{\baselinestretch}{0.9}
    \color{blue}
    \setlength{\parindent}{1em}
    \setlength{\parskip}{0em}
\begin{verbatim}
http://exomol.com/api/?molecule=H2O&datatype=linelist
\end{verbatim}
}
\noindent
returns the details of linelists for the isotopologues of water, omitting partition functions, cooling functions, opacities, etc. Valid values for the \texttt{datatype} parameter are: \texttt{linelist}, \texttt{energylevels}, \texttt{opacity}, \texttt{super} (``Super'' line lists), \texttt{Cp} (heat capacity), \texttt{broadening\_coefficients}, \texttt{coolingfunction}, \texttt{partitionfunction}.

\subsubsection{Accessing the data via the API}\label{sect:data}

There are a number of ways of accessing the data. First, all data sets are available on the ExoMol website (\url{www.exomol.com}) and can be downloaded manually.
Many ExoMol line lists contain in excess of 10 billion transitions. In this
case the file of transitions are generally split into chunks in frequency. These files are compressed
using \texttt{.bz2} format.

However, if multiple datasets are required manual downloads are inefficient. The systematic structure of the ExoMol database filesystem allows for automated downloads from within software or using utilities such as \texttt{wget} and \texttt{curl}.
As described in detail in the ExoMol2016 release, the API structures the ExoMol data resource files as URLs of the form:
{
    \renewcommand{\baselinestretch}{0.9}
    \color{blue}
    \setlength{\parindent}{1em}
    \setlength{\parskip}{0em}
\begin{verbatim}
http://exomol.com/db/<MOLECULE>/<ISOTOPOLOGE>/<DATASET>/<FILENAME>
\end{verbatim}
}
\noindent
where \texttt{<FILENAME>} is structured \texttt{\textquotesingle{}<ISOTOPOLOGUE>\_\_<DATASET>.<DATATYPE>\textquotesingle{}}. For example, the States file of the YYLT line list for $\mathrm{^{31}P^{15}N}$ is obtainable at the URL \url{http://exomol.com/db/PN/31P-15N/YYLT/31P-15N__YYLT.states.bz2}.

Furthermore, each dataset has a \emph{manifest} that lists the data files it is comprised of and their sizes (in bytes). The manifest is a single text file located at the above URL with the filename \texttt{\textquotesingle{}<ISOTOPOLOGUE>\_\_<DATASET>.manifest\textquotesingle{}}. For example, the contents of \url{http://exomol.com/db/PN/31P-15N/YYLT/31P-15N__YYLT.manifest} are:
{
    \renewcommand{\baselinestretch}{0.9}
    \color{blue}
    \setlength{\parindent}{1em}
    \setlength{\parskip}{0em}
\begin{verbatim}
http://exomol.com/db/PN/31P-15N/YYLT/31P-15N__YYLT.states.bz2 142161
http://exomol.com/db/PN/31P-15N/YYLT/31P-15N__YYLT.trans.bz2 1803083
http://exomol.com/db/PN/31P-15N/YYLT/31P-15N__YYLT.pf 175000
http://exomol.com/db/PN/31P-15N/YYLT/31P-15N__YYLT.def 4761
http://exomol.com/db/PN/31P-15N/YYLT/31P-15N__YYLT.manifest 323
\end{verbatim}
}

\section{Moving to higher spectral resolution}

\begin{table}[H]
\tiny
\caption{Molecules of importance for the ExoMol project with published MARVEL datasets.}
\label{tab:marveldata}
\small
\begin{tabular}{lcrrl}
\hline\hline
Molecule&$N_{\rm iso}$&$N_{\rm elec}$&$N_{\rm levels}$&Reference(s)\\
\hline
 H$_2$O &7&1&18~486&\citet{jt454,jt482,jt539,jt576}\\
 \ \ update &1&1& 19~200     &\citet{jt750,jt795}\\
 H$_3^+$&3&1& 652& \citet{13FuSzFa.H3+,13FuSzMa.H3+}\\   
 NH$_3$&1&1&4951& \citet{jt608}\\
 \ \ update&1&1&4936 & \citet{jt784}\\
 C$_2$&1&14&5699& \citet{jt637}\\
 \ \ update & 1&20 & 7087& \citet{jt809}\\
 TiO&1& 9&10~564& \citet{jt672,jt760}\\
 HCCH&1&1&11~213& \citet{jt705}\\ 
 SO$_2$&3&1&15~130& \citet{jt704}\\ 
 H$_2$S&1&1&11~213& \citet{jt718}\\
ZrO&1&10&8088& \citet{jt740}\\
O$_2$&1& 6& 4279& \citet{19FuHoKoSo}\\
NH&1&4&1058& \citet{jt705}\\
CaOH&1&5&1954& \citet{jt791}\\
H$_2$CO&1 & 1& 4841& \citet{jtH2COmarvel}\\
NO& 1&1 & 4106 & \citet{jt686} \\
AlH&2 &2 & 331& \citet{jt732} \\
BeH& 3&2 &1264 & \citet{jt722}\\
CN& 1&10& 7779 &\citet{20SyMc}\\
\hline\hline
\end{tabular}

$N_{\rm iso}$ Number of isotopologues considered;\\
$N_{\rm elec}$ Number of electronic states considered;\\
$N_{\rm levels}$  Number of energy levels extracted: value is for the main isotopologue.\\
\end{table}

Transit spectroscopy of exoplanets has thus far been 
performed at rather low resolution; however, very precise spectroscopic data are
required for high resolution Doppler spectroscopy
\cite{13DeBrSn.exo,13BiDeBr.exo,14BrDeBi.exo}. This has proved to be 
an issue for important species \cite{15HoDeSn.TiO}. Indeed the ExoMol datasets
described above were generally constructed with greater emphasis on completeness
than obtaining very precise transition frequencies \cite{jt716}.
In practice, however, the use of empirical energy levels in the 
States file means that some transition frequencies are indeed reproduced with 
high (experimental) accuracy; however, with the ExoMol2016 data structure it was not
possible to tell how accurately a particular transition was predicted.

Although some work has been done on using ExoMol data to provide the input for high resolution studies
\cite{jt782}, there is a clear need to adapt the database to provide the laboratory data needed for
these studies.
For this reason we have updated the data structure to allow the uncertainty in a
particular transition frequency  to be determined. This is done via uncertainties
in the energy levels which are now specified (optionally) in the States file,
see Table~\ref{t:states:format}. 

To improve the accuracy of the predicted spectra, the ExoMol States files are being
systematically updated using empirically-determined energy levels. The main means
of doing this is via the MARVEL (measured active rotational-vibrational energy levels)
procedure \cite{jt412,12FuCsi.method,jt750}. MARVEL inverts available high resolution spectra for
a given isotopologue to give a list of empirical energy levels with associated
uncertainties. For levels determined by MARVEL these uncertainties are now given
in the States file. Otherwise the (usually much larger) uncertainty arising from
the calculation used to generate the line list is given. By combining the uncertainty
of upper and lower states ($\Delta E_u$, $\Delta E_l$)  using the standard formula 
\begin{equation}
 \Delta E_\nu = \frac{1}{\sqrt{2}}\sqrt{\left((\Delta E_u)^2 + (\Delta E_l)^2\right)}
\end{equation}
gives the uncertainty in the transition wavenumber, $\Delta E_\nu$.

To help improve the accuracy of key line lists we have been running MARVEL
projects on relevant molecules. Table~\ref{tab:marveldata} lists the
astronomically important molecules for which MARVEL studies have been completed.
We note that a number of these studies \cite{jt672,jt705,jt718,jt740,jt764} have
been performed as part of the so-called ORBYTS schools outreach project, see
\citet{jt709} for a discussion of this. We are in the process of working through
all the molecules in the ExoMol database running MARVEL projects for those isotopologues
for which there are enough high accuracy laboratory data available to justify this activity.

\section{Other Future development}

The ExoMol project already maintains extensive molecule by molecule bibliography files. These are stored
on LaTeX's BibTex format and are freely accessible at \url{https://github.com/ExoMol/bib/tree/master/exomol}.
Our aim is to make referencing to the original data as easy and convenient as possible, including automatic generation of the list of references to cite in the appropriate format such as BibTex, Endnote etc. Such services have been developed and offered by HITRAN \citep{20SkGoHi}.

We are in the process of moving the database to a more powerful platform. After the move we plan to offer the
ability to compute  cross sections for a given species, temperature and pressure on the fly.
This will be done using super-lines which greatly reduce the computing time. We will also facilitate the computation of
$k$-tables for a given atmospheric model. 

Finally the move should allow much greater integration with the code Tau-REx   \cite{jt593,jt611}.
Tau-REx is an open source retrieval code for exoplanetary atmospheres which has just undergone a major upgrade
\cite{TauRexIII}. Integration with the  Tau-REx-III upgrade via a graphical interface is currently in progress.

\section{Conclusions}

The ExoMol database presented here is a molecule-by-molecule set of
comprehensive line lists for modelling spectra and other properties of
hot gases. The choice of molecules is dictated by the need to model
the atmospheres of exoplanets and other hot astronomical objects, but
the spectroscopic data have much wider applications than this. We are
still in the process of adding molecules to the database and are
receptive to suggestions of other key species to include.  
In addition we are working on improving the accuracy of the line positions, particularly
for strong lines, enhancing the treatment of pressure broadening, and extending the range of
wavelengths covered into the ultra-violet for molecules where this is considered important.
In addition, we also plan to expand the
database to consider temperature dependent photodissociation.

\section*{Acknowledgments}

We thank the many scientists who have contributed directly or indirectly to the
ExoMol project including Ala'a Azzam, Emma Barton, Bob Barber, Phillip Coles, Lorenzo Lodi, Barry Mant, Andrei Patrascu,
Anatoly Pavlyuchko, Clara Sousa-Silva, Tom Rivlin, Daniel Underwood,  Andrei Yachmenev and Peter Bernath.
This work was supported by the European Research Council (ERC) under the Advanced Investigator ExoMol Project
267219 and the ExoAI project 758892, and STFC through grants ST/H008586/1 and ST/K00333X/1.

\bibliographystyle{model3-num-names}

\end{document}